\documentclass[aps,prx,twocolumn,longbibliography,showpacs,superscriptaddress,floatfix]{revtex4-1}

\usepackage{graphicx}
\usepackage{amsmath,amssymb,bbold,bm,color}
\usepackage{float}
\usepackage{epstopdf}
\usepackage{hyperref}
\usepackage{color}
\usepackage{enumerate}
\usepackage{wasysym}
\usepackage{url}
\usepackage{dsfont}
\usepackage{braket}
\hypersetup{colorlinks=true,linkcolor=blue,citecolor=blue,urlcolor=blue}

\begin{document}

\title{Traversable wormhole and Hawking-Page transition in coupled complex SYK models}

\author{Sharmistha Sahoo}
\email{ssahoo@mail.ubc.ca}
\affiliation{Department of Physics and Astronomy \& Stewart Blusson Quantum Matter Institute, University of British Columbia, Vancouver BC, Canada V6T 1Z4}

\author{\'Etienne Lantagne-Hurtubise}
\email{lantagne@phas.ubc.ca}
\affiliation{Department of Physics and Astronomy \& Stewart Blusson Quantum Matter Institute, University of British Columbia, Vancouver BC, Canada V6T 1Z4}
\affiliation{Kavli Institute for Theoretical Physics, University of California Santa Barbara, CA 93106, USA}

\author{Stephan Plugge}
\affiliation{Department of Physics and Astronomy \& Stewart Blusson Quantum Matter Institute, University of British Columbia, Vancouver BC, Canada V6T 1Z4}

\author{Marcel Franz}
\affiliation{Department of Physics and Astronomy \& Stewart Blusson Quantum Matter Institute, University of British Columbia, Vancouver BC, Canada V6T 1Z4}

\date{\today}

\begin{abstract}
Recent work has shown that coupling two identical Sachdev-Ye-Kitaev (SYK) models can realize a phase of matter that is holographically dual to an \emph{eternal traversable wormhole}. This phase supports revival oscillations between two quantum chaotic systems that can be interpreted as information traversing the wormhole. Here we generalize these ideas to a pair of coupled SYK models with {\em complex} fermions that respect a global U(1) charge symmetry. Such models show richer behavior  than conventional SYK models with Majorana fermions and may be easier to realize experimentally. We consider two different couplings, namely tunneling and charge-conserving two-body interactions, and obtain the corresponding phase diagram using a combination of numerical and analytical techniques. At low temperature we find a charge-neutral gapped phase that supports revival oscillations, with a ground state close to the thermofield double, which we argue is dual to a traversable wormhole. We also find two different gapless non-Fermi liquid phases with tunable charge density  which we interpret as dual to a `large' and `small' charged black hole. The gapped and gapless phases are separated by a first-order phase transition of the Hawking-Page type. Finally, we discuss an SU(2)-symmetric limit of our model that is closely related to proposed realizations of SYK physics with spinful fermions in graphene, and explain its relevance for future experiments on this system.

\end{abstract}

\maketitle

\section{Introduction}

The Sachdev-Ye-Kitaev (SYK) model~\cite{SY1993,Sachdev2015,Kitaev2015,Maldacena2016} has emerged recently as a powerful toy model allowing to glean insight into the exotic behavior of non-Fermi liquids, such as quantum chaos~\cite{Maldacena2016b, Jian2017, Gu2017, Banerjee2017}, holography~\cite{Gross2017,Davison2017,Sarosi2018} and strange metallic transport~\cite{Song2017,Wu2018,Patel2018, Chowdury2018, Cha2020}. The model consists of $N$ Majorana fermions coupled via all-to-all, random Gaussian interactions. It derives its predictive power from the fact that, once averaged over quenched disorder, the model can be solved exactly 
through a large-$N$ saddle-point expansion. Remarkably, at low temperature the model can be shown to be maximally chaotic: its out-of-time-order correlators (OTOCs) exhibit an exponentially growing regime with a Lyapunov exponent that saturates the bound on many-body quantum chaos~\cite{Maldacena2016b}. Such growth indicates fast scrambling of quantum information in the system and underlies the holographic connection between the SYK model and black holes. 
Motivated by these exciting predictions, a number of proposals have emerged for the physical realization of the SYK model and its variants in atomic~\cite{Danshita2017}, optical~\cite{wei2020optical} and solid-state~\cite{Pikulin2017,Alicea2017,Rozali2018} platforms, or using quantum simulators~\cite{Solano2017,Luo2019}. 

Interesting physics also occurs when two identical SYK models are coupled by interactions~\cite{KimKlebanov2019} or tunneling~\cite{maldacenaqi2018}. At low temperature the coupling can drive phase transitions to symmetry-broken states~\cite{KimKlebanov2019} or, remarkably, to a phase holographically dual to an \emph{eternal traversable wormhole}~\cite{maldacenaqi2018} with an AdS$_2$ throat (AdS$_2$ refers to the 1+1-dimensional anti-de Sitter spacetime). This phase enables the transmission of information between two chaotic systems through `revival dynamics'~\cite{Plugge2020,Zhang2020} corresponding, in the gravity interpretation, to sending particles through a wormhole~\cite{Gao2017, Yang2017, Maldacena2018, Bak2018, Gao2019, Marolf2019, Bak2019}. Proposals for the physical realization of such coupled SYK models in condensed matter platforms have also been discussed~\cite{LH2019}.  

A key concept at the heart of the traversable wormhole proposal is the thermofield double (TFD) state. Given two identical copies of a quantum mechanical system, a TFD state is defined as
\begin{equation}\label{e1}
    \ket{ \rm{TFD}_{\tilde\beta}} = \frac{1}{\sqrt{Z_{\tilde\beta}}} \sum_n e^{-{\tilde\beta} E_n/2} \ket{n}_1 \otimes {\ket{\bar{n}}}_2.
\end{equation}
Here $Z_{\tilde\beta}=\sum_{n} e^{-{\tilde\beta} E_{n}}$ is the partition function of a single system at inverse temperature ${\tilde\beta}$, and $| \bar{n} \rangle = |\Theta n\rangle$ where $\Theta$ is an anti-unitary symmetry of the Hamiltonian.
The TFD is a purification of a Gibbs state at inverse temperature $\tilde\beta$: it is an entangled state of the two copies, such that tracing over either copy recovers the thermal density matrix for the other. As such, the TFD can be used as a resource state to study thermal properties in quantum simulators~\cite{Hsieh2019}. It also obeys an inverted version of time-translation invariance which enables accessing OTOCs using ordinary time-ordered measurements~\cite{LH2019} and teleporting states or operators between the two subsystems~\cite{Brown2019, Gao2019b}. Recent work has shown how to construct model Hamiltonians with a ground state close to a TFD~\cite{maldacenaqi2018, Cottrell2019} by coupling identical systems, the Maldacena-Qi (MQ) wormhole model~\cite{maldacenaqi2018, Garcia-Garcia2019, alet2020entanglement} being an example of such a construction. 

Variants of the SYK model built from ordinary complex fermions, rather than real Majorana fermions, have also been studied and exhibit similar properties~\cite{Sachdev2015, Gu2020}. The main difference between the complex and Majorana SYK models is that the former has a conserved U(1) charge. Importantly, these complex SYK variants, henceforth abbreviated as cSYK, might be easier to realize in condensed matter systems~\cite{Chen2018,Altland2019}, where Majorana fermions are notoriously difficult to obtain and control. Further, a number of experimental probes such as spectroscopy~\cite{Chen2018, Gnezdilov2018}, electrical conductance~\cite{Can2019} and thermopower~\cite{Kruchkov2020} have recently been proposed to identify signatures of complex SYK models. 

\begin{figure}
    \centering
    \includegraphics[width=0.95\columnwidth]{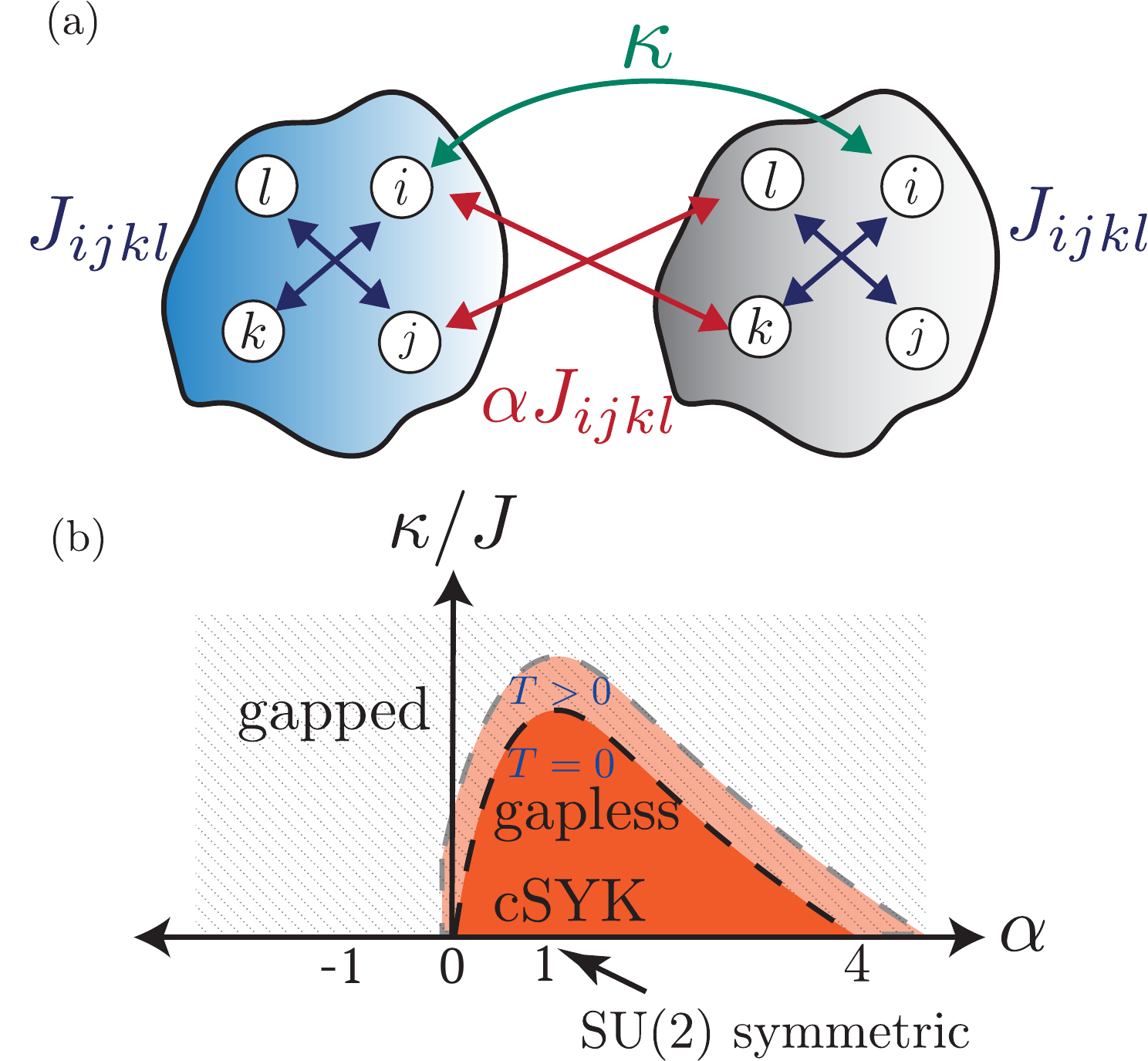}
    \caption{(a) Illustration of the coupling terms in our model. (b) Phase diagram of the model at charge neutrality, for zero and low temperatures. The dashed line indicates a first-order phase transition between a gapless non-Fermi liquid phase and a gapped phase supporting revival oscillations. The $\alpha=0$ line is the generalization of the MQ wormhole model~\cite{maldacenaqi2018} to complex fermions, while the $\alpha=1$ line with SU(2) symmetric interactions maps to the graphene flake proposal of Ref.~\cite{Chen2018}.
    }
    \label{fig:modeldiagram}
\end{figure}

In this work we investigate the physics of \emph{coupled} complex SYK models using a combination of analytical arguments, exact diagonalization for small $N$ and saddle-point solutions for large $N$. We consider two types of couplings: a simple tunneling term with strength $\kappa$, and random two-body interactions with strength $\alpha$  that conserve charge in each system separately. The form of the couplings, illustrated in Fig.~\ref{fig:modeldiagram} (a), is partly motivated by their natural connection to the disordered graphene flake proposal of Ref.~[\onlinecite{Chen2018}].

We first show that coupling identical cSYK models with the tunneling term leads to similar physics as the MQ wormhole  model~\cite{maldacenaqi2018}: at low temperature the system is gapped, with a charge-neutral ground state close to a TFD. The correct definition of the TFD state in the presence of a U(1) symmetry is however subtle and we explain it in detail. We obtain the finite-temperature phase diagram of the model, showing that the gapped phase with a TFD ground state is separated from a gapless cSYK phase at high temperature by a first-order transition line ending at a critical point. Further, we investigate the dynamics of the system and find that two-point correlations between the two subsystems decay as a power-law in the high-temperature cSYK phase, but show periodic revivals in the gapped phase~\cite{Plugge2020,Zhang2020}. These results lead us to conjecture that the gapped phase of this model, similarly to its Majorana counterpart~[\onlinecite{maldacenaqi2018}], is holographically dual to a traversable wormhole.
The first-order phase transition between the wormhole phase with a TFD ground state and the charged black hole (cSYK) phase can thus be interpreted as a Hawking-Page type transition~\cite{Hawking1983}.

We then consider two-body interactions with strength $\alpha$ that conserve charge in each system separately -- leading to a U(1)$\otimes$U(1) symmetry group. At charge neutrality we find two low-temperature phases depending on the interaction strength $\alpha$. For $\alpha < 0$ or $\alpha > 4$, we observe a gapped phase with spontaneous symmetry-breaking from U(1)$\otimes$U(1) down to U(1), while for $0 \leq \alpha \leq 4$ we obtain a gapless non-Fermi liquid phase with tunable charge density and properties similar to the cSYK phase.
Combining both types of couplings we find the phase diagram illustrated in Fig.~\ref{fig:modeldiagram} (b) which contains a dome-shaped gapless cSYK phase, separated from the surrounding gapped phase by a zero-temperature first-order phase transition. Surprisingly, away from charge neutrality we uncover another first-order phase transition to a \emph{different} gapless non-Fermi liquid describing a smaller black hole with half of its degrees of freedom gapped out. We finally discuss in detail the $\alpha=1$ limit of our model, featuring SU(2)-invariant interactions, which is directly relevant to the proposed graphene flake realization of the cSYK model~\cite{Chen2018}. Using this new connection we revisit the results of Ref.~[\onlinecite{Chen2018}] and argue that the irregularly-shaped graphene flake model for \emph{weak} applied magnetic field admits a cSYK non-Fermi liquid phase. 

The rest of this paper is organized as follows. In Sec.~\ref{sec:TFD}, we review  properties of the TFD and show how to define it for a pair of identical complex SYK models with U(1) symmetry. In Sec.~\ref{sec:model}, we couple the two systems via the tunneling term and discuss the resulting gapped phase, which we conjecture is dual to a wormhole in the gravity description. In Sec.~\ref{sec:alphamodel}, we add interaction terms between the two systems and explore the resulting phase diagram. In Sec.~\ref{sec:graphene} we provide a connection of our model to the graphene flake proposal~\cite{Chen2018} with SU(2)-symmetric interactions, and discuss its relevance for future experiments on this system. More technical contributions, including details of the TFD construction and solutions of the saddle-point equations, in both imaginary and real time, are presented in the Appendices.

\section{\label{sec:TFD}Thermofield double state construction}
Consider two identical copies of a quantum system described by Hamiltonians $ H_1 = H_2$, with eigenstates $\ket{n}_1$ and $\ket{n}_2$ of a common eigenenergy $E_n$. A TFD is an entangled state of the two copies defined by Eq.\ \eqref{e1}. Such a definition represents a one-parameter family of states which can be generated from
\begin{equation}\label{e2}
\ket{ \rm{TFD}_{\tilde{\beta}} } = e^{-\tilde{\beta} (H_{1}+H_{2})/4} \ket{I}
\end{equation}
where $\ket{ I } = \ket{ \rm{TFD_0}}$.
(We denote the effective inverse temperature of the TFD state as $\tilde{\beta}$ to differentiate it from the physical temperature $T=1/\beta$ of the system). The advantage of the representation in Eq.~\eqref{e2} is that $\ket{I}$ is a maximally entangled state between the two copies, and is thus usually simple to write down. 

The TFD state has a few important properties which follow from its definition, Eq.~\eqref{e1}. First, it is a purification of the thermal density matrix, such that
\begin{equation}
    \braket{ V_a }_{\tilde{\beta}}  = \frac{1}{Z_{\tilde{\beta}}} {\rm Tr} \left[ V e^{-\tilde{\beta} H} \right]
    \label{eq:TFD_thermal}
\end{equation}
for any operator $V_a$ ($a=1,2$) that acts only on one subsystem. Here the expectation value is taken in the state $\ket{\rm{TFD}_{\tilde{\beta}} }$, and the trace is over the Hilbert space of a single subsystem. Although the TFD is \emph{not} an eigenstate of the full system's Hamiltonian $H_1 + H_2$, it is an eigenstate of the difference
\begin{equation}
    \left( H_1 - H_2 \right) \ket{\rm{TFD}_{\tilde{\beta}}} = 0
\end{equation}
which implies that two-point correlation functions respect an inverted version of time-translation invariance,
\begin{equation}
    \braket{ V_1(t) W_2(t') }_{\tilde{\beta}}  = \braket{ V_1(t+t'') W_2(t'-t'') }_{\tilde{\beta}},
\end{equation}
for any operators $V_1$ and $W_2$. This property has an analog in the gravity context, where the TFD is used to describe traversable wormholes and time effectively flows in opposite directions on its two sides~\cite{Yang2017, maldacenaqi2018}.

Our goal in this section is to construct the TFD state when each subsystem is described by a cSYK model. As we shall see this construction entails a subtlety: Whereas in the Majorana SYK model the state $\ket{ \bar{n} }$ can be chosen as equal to $\ket{ n}$ (up to a phase), this is not the case for cSYK where $\ket{ \bar{n} }$ carries a different charge quantum number.

\subsection{Complex SYK model}
The complex SYK model is the charge conserving variant of the Majorana SYK model. It is described by the Hamiltonian
\begin{align}
H= \sum_{i,j,k,l=1}^{N} J_{ij;kl} c^{\dagger}_{i} c^{\dagger}_{j} c_{k} c_{l} - \mu \sum_{i}c^{\dagger}_{i} c_{i}
\label{eq:cSYK_Hamiltonian}
\end{align} 
where the $c_{i}$, $c_i^\dagger$ are $N$ fermionic operators satisfying $\{ c_i, c_j^\dagger \} = \delta_{ij}$. The coefficients $J_{ij;kl}$ are complex Gaussian random numbers with 
\begin{equation}
    \overline{J_{ij;kl}} = 0 \quad , \quad \overline{ |J_{ij;kl}|^2 } = \frac{J^2}{8 N^3}
\end{equation}
and satisfy the symmetry constraints
\begin{equation}
J_{ij;kl}=-J_{ji;kl} = -J_{ij;lk} = J_{lk;ji}^{\ast}
\end{equation}
imposed by the fermionic commutation relations. The model has a global U(1) symmetry, $c_j \rightarrow c_j e^{-i\phi}$ that expresses the conservation of the total charge 
\begin{equation}
Q = \sum_{i = 1}^{N} (c_{i}^{\dagger}c_{i}-\frac{1}{2}).
\end{equation}
A detailed discussion of the cSYK model and its physical properties can be found in Refs.~[\onlinecite{Sachdev2015,FuSachdev2016,Gu2020}].

Additionally, at charge neutrality ($\mu=0$) the model has an anti-unitary particle-hole symmetry if we constrain the tensor of couplings $J_{ij;kl}$ to be fully antisymmetric. This anti-unitary symmetry is generated by
\begin{align} P = \prod_{a=1}^N (c_{a}+c_{a}^\dagger) \mathcal{K},
\end{align}
with $\mathcal{K}$ the complex conjugation operator, and transforms $c_j \leftrightarrow c_j^{\dagger}$ up to a sign that depends on $N$. One can check that $P^{-1} H P = H$ and
\begin{equation}
P^{2} = (-1)^{N(N-1)/2}.
\end{equation} 
Such a symmetry has been discussed in Refs.~[\onlinecite{Gu2020, FuSachdev2016, Youetal2017,behrends2019symmetry}] and is useful to simplify calculations. Ref.~[\onlinecite{FuSachdev2016}] implements it by including additional hopping terms to cancel out the unwanted interaction terms with repeated indices. Here we simply consider fully anti-symmetric $J_{ij;kl}$, and use the anti-unitary symmetry $P$ to define the states $\ket{\bar{n}}$ that are required to construct the TFD state.  

Leveraging the U(1) symmetry, the Hamiltonian Eq.~\eqref{eq:cSYK_Hamiltonian} can be block diagonalized in symmetry sectors labeled by the eigenvalues $q$ of the charge operator $Q$. Since $P^{-1} Q P = - Q$, at charge neutrality the model has a two-fold spectral degeneracy guaranteed by the mapping between states in the $q$ and $-q$ sectors. For even $N$ there is a special zero-charge ($q=0$) sector which maps onto itself under $P$; this enforces a two-fold Kramers degeneracy when $P^{2}=-1$, for $N~\text{mod}~4 = 2$ (see Table~\ref{tab:deg}). 

\subsection{Thermofield double state for complex SYK}
As discussed above, the eigenstates $\ket{n_q}$ of the cSYK model, in the presence of the anti-unitary symmetry $P$, form particle-hole pairs in charge sectors $\pm q$. We thus define the TFD as a zero-charge state with
 \begin{equation} \label{TFD} 
 |{\rm TFD}_{\tilde{\beta}}\rangle = \frac{1}{\sqrt{Z_{\tilde{\beta}}}} \sum_{q=-N/2}^{N/2} \sum_{n_q}  e^{-\tilde{\beta} E_{n}/2} \ket{n_q}_1 \otimes \ket{ \bar{n}_{-q} }_2
 \end{equation} 
 where the state $\ket{ \bar{n}_{-q} }$ is equivalent to $ P \ket{ n_q }$ up to a phase which depends only on the symmetry label $q$. (An arbitrary phase $\theta_n$ is not allowed as it cannot be absorbed by a gauge transformation in $\ket{n}$ and $\ket{\overline{n}}$.) This phase can be fixed most easily by choosing a specific infinite-temperature ($\tilde{\beta}=0$) TFD state $\ket{I}$ and identifying it with $\ket{\rm{TFD}_{0}}$. The definition in Eq.~\eqref{e2} then automatically fixes the same phases for the finite-temperature TFD states. The state $\ket{I}$ must be maximally entangled, as it is the purification of the infinite-temperature density matrix.
This motivates the choice of a product of $N$ Bell pairs between the two systems~\footnote{Other choices of maximally-entangled states are in principle possible. Our choice of this particular Bell state is motivated by the physical coupling introduced in Sec.~\ref{sec:model}, and is reflected in the phase factors appearing in the definition of the anti-unitary symmetry, Eq.\eqref{eq:theta_definition}.},
\begin{equation}
|I_\phi\rangle = \prod_{i=1}^N \frac{1}{\sqrt{2}} \left( |1\rangle_{1}|0\rangle_{2} - e^{-i \phi} |0\rangle_{1}|1\rangle_{2} \right)_{i}.
\label{eq:definition_I}
\end{equation}
In Appendix~\ref{appendix:TFD} we explicitly show that the state in Eq.~(\ref{eq:definition_I}) is equivalent to
\begin{equation}\label{iTFD}
\ket{ \rm{TFD}_0 } \equiv \frac{1}{2^{N/2}}\sum_{q} \sum_{n_q} \ket{n_q}_1 \otimes \Theta  \ket{ n_q }_2
\end{equation}
with the anti-unitary symmetry defined as
\begin{equation}\label{eq:theta_definition}
\Theta = e^{-\frac{i\eta \pi \Gamma}{4}} e^{-iq(\phi-\frac{\pi}{2})} P.
\end{equation}
Here $q$ and $\Gamma = (-1)^{q+\frac{N}{2}}$ are the charge and fermion parity, respectively, of the state $\ket{ n_q}$ and $\eta = \pm1$ is a sign which depends on the total number $N$ of fermions through $P^{-1} c_{i1} P = \eta c_{i1}^{\dagger}$. The parity-dependent phase factor has been discussed for the Majorana version of the TFD state~\cite{Garcia-Garcia2019}. The charge-dependent phase is required to cancel the minus signs appearing when fermionic creation or annihilation operators are taken across eigenstates of subsystem $1$ to act on subsystem $2$. Note that the TFD for a bosonic Hamiltonian would not have these phases.

\section{\label{sec:model} Coupling $c$SYK models with tunneling term}

In this section we introduce a coupling between the two cSYK Hamiltonians which results in a ground state close to $\ket{ \rm{TFD}_{\tilde{\beta}} }$. Consider the infinite-temperature TFD state $\ket{I_\phi}$ defined in Eq.~(\ref{eq:definition_I}). This is the ground state of a simple tunneling term
\begin{equation}
 K= \sum_{i} \kappa \left( e^{i \phi} c_{i1}^{\dagger} c_{i2} +  e^{-i \phi} c_{i2}^{\dagger} c_{i1} \right)
\end{equation}
with real $\kappa$ and $\phi$. Further, it is clear that the zero-temperature TFD state $\ket{ \rm{TFD}_{\infty}} = \ket{0}_1 \otimes \ket{\overline{0}}_2$ is the exact ground state of two decoupled identical cSYK models. We thus consider the following model,
\begin{align} H_{\kappa} =& \sum_{ij;kl} J_{ij;kl} \sum_{a=1,2} c_{ia}^{\dagger}c_{ja}^{\dagger} c_{ka} c_{la} - \mu \sum_{i,a }c^{\dagger}_{ia} c_{ia} + K ,
\label{eq:model_tunneling}
\end{align} 
where the coupling constants $J_{ij;kl}$ are identical in systems 1 and 2. For large coupling $\kappa /J \gg 1$ the ground state of this Hamiltonian is the Bell state $\ket{I_\phi} = \ket{\rm{TFD}_0}$, whereas for $\kappa/J=0$ it is simply the product state $\ket{\rm{TFD}_{ \infty } }$. As we shall see below using numerical exact diagonalization, the model admits a ground state close to $\ket{ \rm{TFD}_{\tilde{\beta}} }$ for all $\kappa/J$, with a parameter $\tilde{\beta}$ that is a monotonically decreasing function of $\kappa/J$.
Note that a gauge transformation on the fermion operators in either subsystem can absorb the phase $\phi$. Hence in the following we consider, without loss of generality, a purely imaginary tunneling term with $\phi=\pi/2$, corresponding to the simplest case where the charge-dependent phases in Eq.~\eqref{eq:theta_definition} disappear. 

\begin{table}[]
\centering
\begin{tabular}{c|c|c|c}
  $N$ mod 4 & 0 & 1 and 3 & 2\\
  \hline
  \hline
  single cSYK    & 1 & 2 & 2\\
  two decoupled cSYKs & 1 & 4 (1,2,1) & 4 \\
  tunneling ($\kappa > 0$) & 1 & 1 & 1 \\
  interaction ($\kappa=0$, $ 0 < \alpha < \alpha_c$) & 1 & 2 (1,0,1) & 1  \\
interaction ($\kappa=0$, $\alpha<0$ or $\alpha > \alpha_c$) & 1 & 2 (0,2,0) & 1
\end{tabular}
 \caption{Ground state degeneracy of two identical cSYK models at charge neutrality $\mu=0$, for various system sizes and model parameters $\kappa, \alpha$. Ground states are in the $q=0$ symmetry sector unless indicated in brackets, which show the number of ground states in the $q= (-1,0,1)$ sectors. The critical value $\alpha_c \sim 4$.}
 \label{tab:deg}
\end{table}

The symmetries and the ground state degeneracy of the model, deduced by simple arguments and verified using exact diagonalization, are summarized in Table~\ref{tab:deg}. For $\kappa=0$ we have two decoupled cSYK models, where the charge is separately conserved in systems $1$ and $2$ (U(1)$\otimes$U(1) symmetry). There are two anti-unitary symmetries $P_{1} = P \otimes \mathbf{1}$ and $P_{2} = \mathbf{1} \otimes P$, where $P$ acts in one subsystem. The ground state degeneracy is the product of that of each SYK model, which is unique for $N \; \text{mod} \; 4 = 0$ and doubly degenerate otherwise. (The ground states of a single SYK model are always found in the $q=0$ sector (for even $N$) or the $q=\pm 1/2$ sectors (for odd $N$)). For even $N$, the ground states of the coupled system are in the $q=0$ sector, while for odd $N$ they are distributed in the $q=0, \pm 1$ sectors as shown in Table~\ref{tab:deg}. The tunneling term $\kappa$ breaks the charge conservation in each system down to the total U(1) charge conservation, and also breaks the anti-unitary symmetries $P_1$ and $P_2$ -- thus the ground state for $\kappa > 0$ is unique. Finally, there is a discrete mirror symmetry that exchanges fermion operators between systems $1$ and $2$. For $\phi =  \pi/2$ this symmetry transforms $c_{i1} \rightarrow c_{i2}$ and $c_{i2} \rightarrow -c_{i1}$, and constrains the two-point correlators of the system as discussed in Sec.~\ref{sec:largeN}.

\subsection{Exact diagonalization: TFD ground state  \label{ED}}

\begin{figure}
\includegraphics[width=\columnwidth]{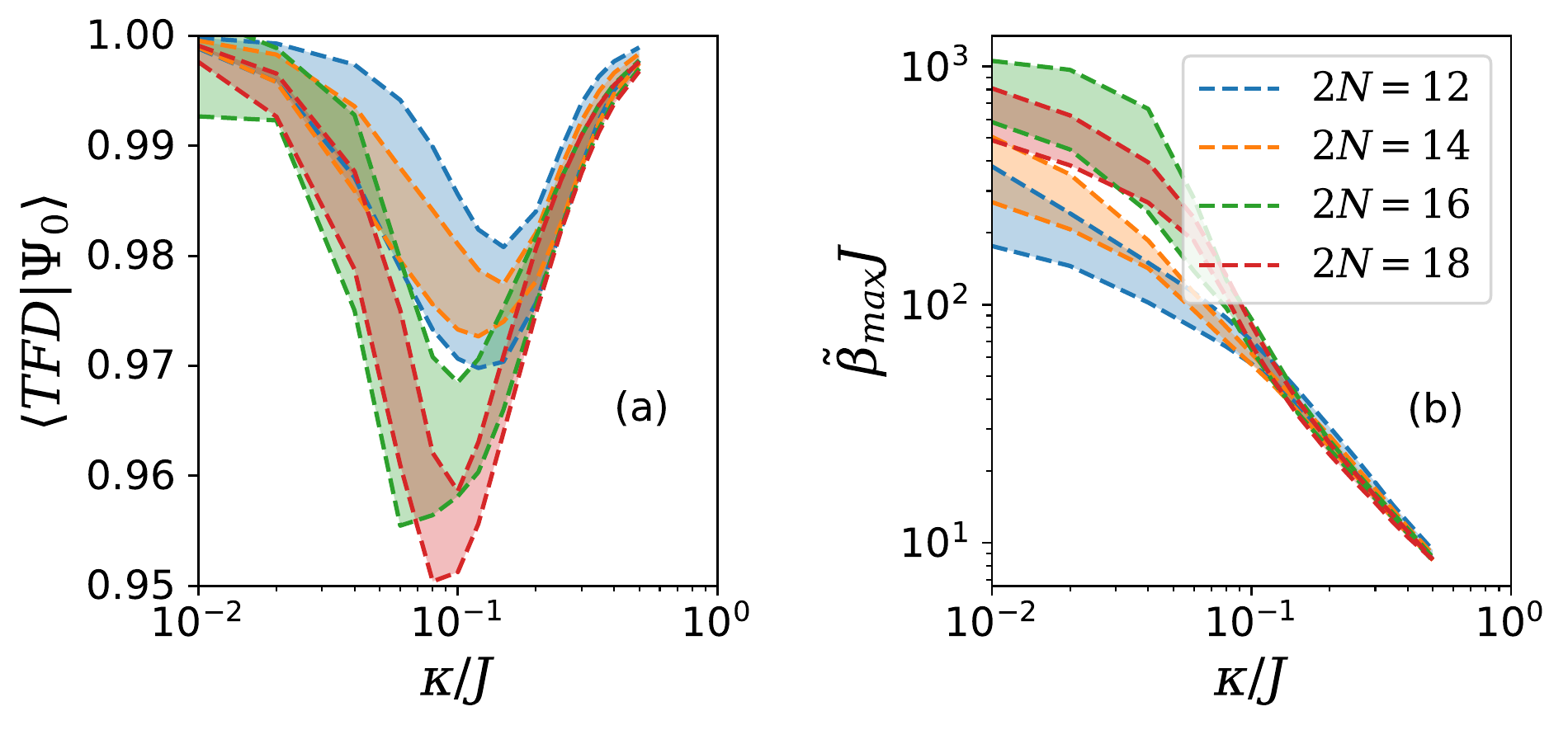}
\caption{Exact diagonalization results for $2N$ up to 18 at charge neutrality $\mu=0$. The shaded area corresponds to the standard deviation obtained from $12$ independent disorder realizations. (a) Overlap between the ground state $\ket{\Psi_0}$ of the coupled cSYK models, Eq.~(\ref{eq:model_tunneling}) and the best-fit TFD state as a function of $\kappa/J$. (b) Effective inverse temperature $\tilde{\beta}_{\rm max}$ of the best-fit $\rm{TFD}$ state.
\label{fig:ED_zeroalpha} }
\end{figure}

We first perform an exact diagonalization study of the model, Eq.~(\ref{eq:model_tunneling}), to confirm that it admits a ground state close to a TFD. To do this we construct the family of TFD states with parameter $\tilde{\beta}$ from the eigenstates of a single SYK model using the definition, Eqs.~(\ref{TFD})-(\ref{eq:theta_definition}). We then compute the overlap of this family of TFD states with the numerical ground state of the coupled model, Eq.~(\ref{eq:model_tunneling}) and select the TFD with the parameter $\tilde{\beta}$ that maximizes the overlap. As shown in Fig.~\ref{fig:ED_zeroalpha} this best-fit overlap is always close to $1$, with a minimum of $\sim 0.96$ at a value $\kappa/J \sim 0.1$  which roughly corresponds to the end of the finite temperature first-order transition seen in the large-$N$ calculation (see Sec.~\ref{sec:largeN}). The parameter $\tilde{\beta}$ characterizing the best-fit TFD is monotonically decreasing with $\kappa$. In the gravity interpretation of the MQ model the parameter $\tilde{\beta}$ is proportional to the length of the wormhole, or equivalently to the period of the revival oscillations between the two sides~\cite{maldacenaqi2018, Plugge2020}.

\subsection{Large $N$ saddle point solution\label{sec:largeN}}

We now derive the large-$N$ saddle point equations of the model Eq.~(\ref{eq:model_tunneling}) in imaginary (Euclidean) time $\tau$, and solve them numerically to investigate its properties 
in the thermodynamic limit. The partition function of the system at inverse physical temperature $\beta$ is given by
\begin{align} \mathcal{Z} = \int \prod_{i} \mathcal{D} c_{i a} \mathcal{D} c^{\dagger}_{i a} e^{-\int_0^\beta d\tau (\sum_{i,a=1,2} c_{i a}^{\dagger}(\tau) \partial_{\tau} c_{i a}(\tau) + H_{\kappa})}. \end{align} 
Upon disorder averaging (keeping only replica-diagonal terms) and integrating out the fermion fields we arrive at the following effective action (see Appendix~\ref{app:largeN}),
\begin{align}
-\frac{S[G,\Sigma]}{N} &= \ln {\rm Det} M + \sum_{a,b}  \int_{\tau, \tau'} \Big[ \Sigma_{ab}(\tau,\tau') G_{ba}(\tau',\tau) \nonumber \\
&+ \frac{J^{2}}{4}G_{ab}^2(\tau,\tau') G^2_{ba}(\tau',\tau) \Big]
\label{eq:effective_action}
\end{align}
where $G_{ab}(\tau,\tau')= \frac{1}{N} \langle \mathcal{T}\sum_{i} c_{i a}(\tau) c^{\dagger}_{i b}(\tau') \rangle$  is the averaged time-ordered correlator at the saddle point and $\Sigma_{ab}(\tau, \tau')$ are Lagrange multipliers which can be interpreted as fermion self-energies.  The matrix $M = \bigoplus_n M(i \omega_n)$ with
$M_{ab}(i \omega_n) = (- i \omega_n - \mu)\delta_{ab} + i \kappa \epsilon_{ab} - \Sigma_{ab}(i\omega_n)$
results from performing the integral over complex Grassman fields. 

Varying the effective action Eq.~(\ref{eq:effective_action}) with respect to $G_{ab}$ and $\Sigma_{ab}$  leads to the large-$N$ saddle-point equations. Using the time-translation symmetry $G_{ab}(\tau,\tau') = G_{ab}(\tau - \tau')$ and the mirror symmetry which enforces $G_{11}(\tau) = G_{22}(\tau)$ and $G_{12}(\tau) = -G_{21}(\tau)$, the saddle-point equations can be reduced to the form
\begin{align} 
\label{SDeqn}
G_{11}(i\omega_{n}) &= \frac{-i \omega_{n}-\mu -\Sigma_{11}(i \omega_n)}{D(i\omega_{n})}~,
\nonumber\\
G_{12}(i\omega_{n}) &= \frac{-i \kappa + \Sigma_{12}(i \omega_n)}{D(i\omega_{n})}~,\nonumber\\ 
\Sigma_{11}(\tau) &= - J^2 G_{11}^{2}(\tau) G_{11}(-\tau)~,\nonumber\\
\Sigma_{12}(\tau) &= J^2 G_{12}^{2}(\tau) G_{12}(-\tau)~,
\end{align}
where $\omega_{n} = (2n+1)\frac{\pi}{\beta}$ are fermionic Matsubara frequencies, 
\begin{equation}
  D(i\omega_{n}) = (-i \omega_{n}-\mu - \Sigma_{11})^2 + (i \kappa - \Sigma_{12})^2,
\end{equation}
 and the parameters $J$, $\mu$ and $\kappa$ are real. When $\kappa = 0$ and thus $G_{12}=0$, these equations reduce to the usual cSYK model, which can be solved in the low-frequency or long-time limit by appealing to an emergent conformal invariance. The zero-temperature result is~\cite{Sachdev2015}
\begin{equation}
    G_{11}(\tau) = 
    \begin{cases}
    b\tau^{-1/2} & \tau \gg J^{-1} \\
    -b e^{-2 \pi {\cal E}} |\tau|^{-1/2} & -\tau \gg J^{-1} 
    \end{cases}
\label{eq:correlator_SYK}
\end{equation}
where $b=1/(4\pi J^{2})^{1/4}$. The `twist' parameter ${\cal E}$, which leads to a spectral asymmetry in the frequency domain,
comes about because the chemical potential $\mu \neq 0$ breaks the particle-hole symmetry of the problem. The resulting U(1) charge density $\mathcal{Q} = \braket{Q}/N$ is related to ${\cal E}$ by~\cite{Georges2001,Sachdev2015} 
\begin{equation}
\label{eq:Qvsepsilon}
\mathcal{Q} = \frac{1}{4} \left[ 1 - \tanh(2\pi {\cal E})\right] - \frac{1}{\pi} \tan^{-1}(e^{2 \pi {\cal E}}).
\end{equation}
The asymmetry parameter ${\cal E}$ is also related to the thermodynamic quantity
\begin{equation}
{\cal E} = \frac{1}{2\pi} \frac{\partial {\cal S}_{0} }{\partial \mathcal{Q}}
\label{eq:S0Q}
\end{equation}
where $\mathcal{S}_{0}$ is the residual zero-temperature entropy density of the cSYK model. In the holographic picture ${\cal E}$ describes the electric field near the charged AdS$_2$ black hole horizon, and respects Eq.~(\ref{eq:S0Q}) with $\mathcal{S}_0$ the Bekenstein-Hawking entropy density of the horizon~\cite{Sachdev2015, Sachdev2019}. 

\begin{figure*}
\begin{center}
\includegraphics[width=\textwidth]{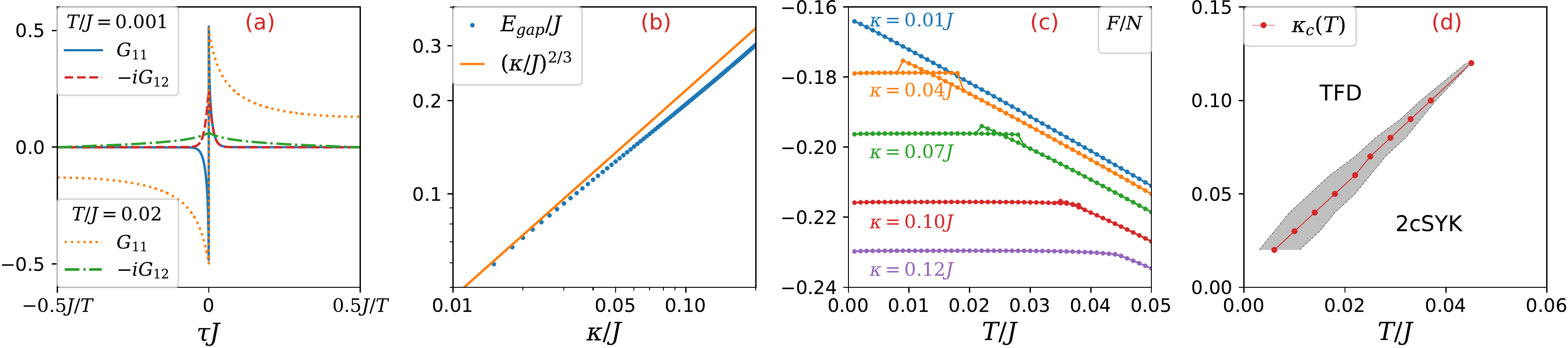}
\end{center}
\caption{\label{gapscalingcSYK}  (a) Imaginary time correlators $G_{11}(\tau)$ and $G_{12}(\tau)$ for $\kappa/J = 0.03$ and temperatures $T/J = 0.001$ and $0.02$. While the solution for $G_{11}$ at low temperature decays exponentially at long times $1 \ll \tau J \ll J/T$, at higher temperature it instead follows a power-law $G_{11} \sim \text{sgn}(\tau) |\tau|^{-1/2}$ indicative of SYK behavior. (b) Energy gap extracted from the exponential decay of correlators $G_{11}(\tau)$ and $G_{12}(\tau)$ as a function of $\kappa$ at low temperature, $T/J = 10^{-4}$. (c) Free energy density as we decrease temperature from $T/J=0.05$ to low temperature, and then increase back to $T/J=0.05$. The hysteresis indicates a first-order phase transition when the two solutions cross. (d) The transition temperature $T_{c}$ as a function of the tunneling strength $\kappa/J$. The shaded region indicates the parameter range where the two phases coincide in (c).}
\end{figure*}

Let us first consider the model at charge neutrality, $\mu=0$. When $\kappa$ (and thus $G_{12}$) is non-zero, an exact solution of the saddle-point equations \eqref{SDeqn} in the low-energy limit cannot be obtained. Instead we solve them numerically by iterating until a self-consistent solution is found, choosing the initial seeds for the iteration to be the  non-interacting solution $G_{11}(\tau)=\frac{1}{2}\mbox{sgn}(\tau)$, and $G_{12} = i\epsilon$ with small $\epsilon$. We find that only the real and imaginary part of $G_{11}$ and $G_{12}$, respectively, are non-zero. For $\kappa=0$ we recover the conformal result Eq.~\eqref{eq:correlator_SYK} for long times and low temperatures $ J^{-1} \ll \tau \ll\beta$. As shown in Fig.~\ref{gapscalingcSYK}(a), when turning on a small coupling $\kappa/ J= 0.03$, a gap opens at low temperature,
as indicated by the exponential decay of the correlators $G_{11}$ and $G_{12}$. For high temperatures 
the correlators instead decay as a power law. In Fig.~\ref{gapscalingcSYK}(b) we show that the energy gap extracted from the exponential decay of $G_{ab}(\tau)$ at low temperature $T/J = 10^{-4}$ scales as $\sim (\frac{\kappa}{J})^{2/3}$ for $\kappa/J \ll 1$. This is consistent with the scaling  of the analogous MQ model~\cite{maldacenaqi2018}.

The MQ model also exhibits a first-order phase transition at finite temperature for small values of $\kappa$. In the gravity context this transition was interpreted~\cite{maldacenaqi2018,maldacena2019} as a Hawking-Page transition~\cite{Hawking1983} because it separates a stable AdS$_2$ black hole at high temperature from a low-temperature phase (the wormhole) which appears thermal for an observer having access to only one subsystem. This is also manifest in the wormhole phase admitting a TFD ground state (see Eq.~\ref{eq:TFD_thermal}).
Such a transition can be identified from the thermodynamics of our complex fermion model. The free energy $F = -T \ln \mathcal{Z}$ is obtained by substituting the saddle point solutions in the action,

\begin{align} \label{eq:free_energy}
\frac{F}{N} &=-T \Big[ 2 \ln 2 + \sum_{\omega_{n}} \ln \left( \frac{D(i\omega_{n})}{(i\omega_{n})^2} \right) \\ & +\frac{3}{2} \sum_{\omega_{n}} \left( \Sigma_{11}(i\omega_{n}) G_{11}(i\omega_{n})- \Sigma_{12}(i\omega_{n}) G_{12}(i\omega_{n})\right) \Big] \nonumber.
\end{align} 
\begin{figure}
\includegraphics[width=0.95\columnwidth]{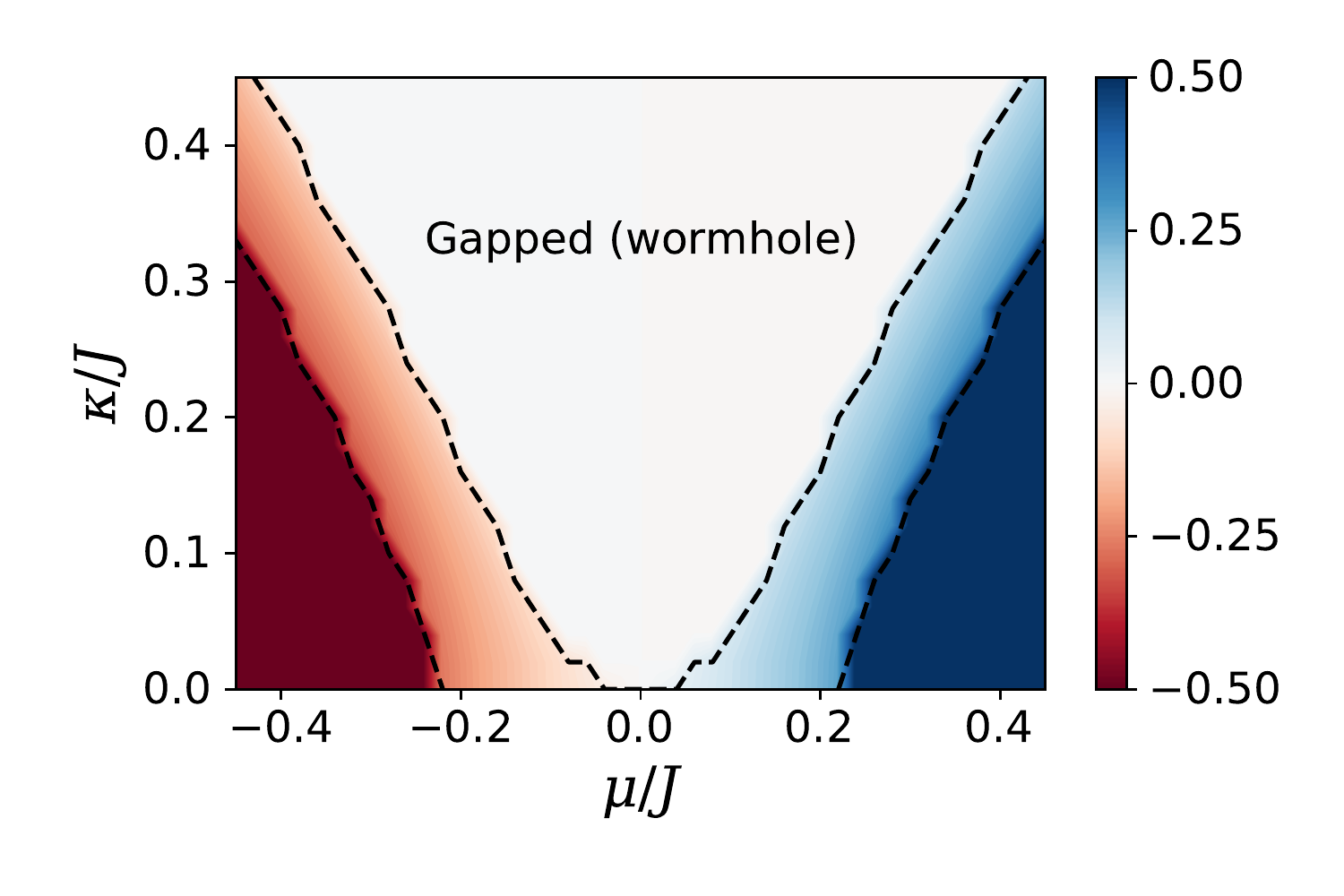}
    \caption{Charge density $\mathcal{Q}$ as a function of $\kappa$ and $\mu$, extracted numerically from the saddle-point solutions through Eq.~(\ref{eq:charge_zerotime}) at low temperature $T/J = 0.002$. The gapped wormhole phase is stable to inclusion of a finite chemical potential $\mu$ and remains charge neutral throughout. For larger $|\mu|$ the system transitions to a gapless phase with tunable charge density, then finally to a gapped polarized phase with $\mathcal{Q}=\pm 1/2$. The dashed lines denote first-order phase transitions.}
    \label{fig:munonzero_MQ}
\end{figure}
Here we regularized the free energy using its value for $2N$ non-interacting complex fermions, $2 \sum_{n} \ln(i\omega_{n}) = 2 \ln 2$, to cancel out divergences at large frequencies in the numerical evaluation of $D(i \omega_n)$. In Fig.~\ref{gapscalingcSYK}c we present the free energy density numerically obtained by sweeping from high to low temperatures (starting each iteration with the converged solution at the previous temperature), and vice versa. For high temperatures the gapless cSYK solution is favored, whereas for low temperatures the gapped solution prevails. These two phases can be identified from the temperature dependence of the free energy: a constant negative slope $S= -\partial F/\partial T$ at low temperatures indicates an SYK phase with residual entropy $S_0$, while the gapped wormhole phase with a unique ground state shows zero slope. We obtain a clear hysteresis between the two solutions indicating a first-order phase transition. The transition temperature $T_c$, identified with the crossing point of the gapped and gapless free energies in Fig.~\ref{gapscalingcSYK}(c), increases monotonously with $\kappa/J$ until the phase transition line terminates at a critical point, as shown in Fig.~\ref{gapscalingcSYK}(d). 
\begin{figure*}
	\includegraphics[width=1.0\textwidth]{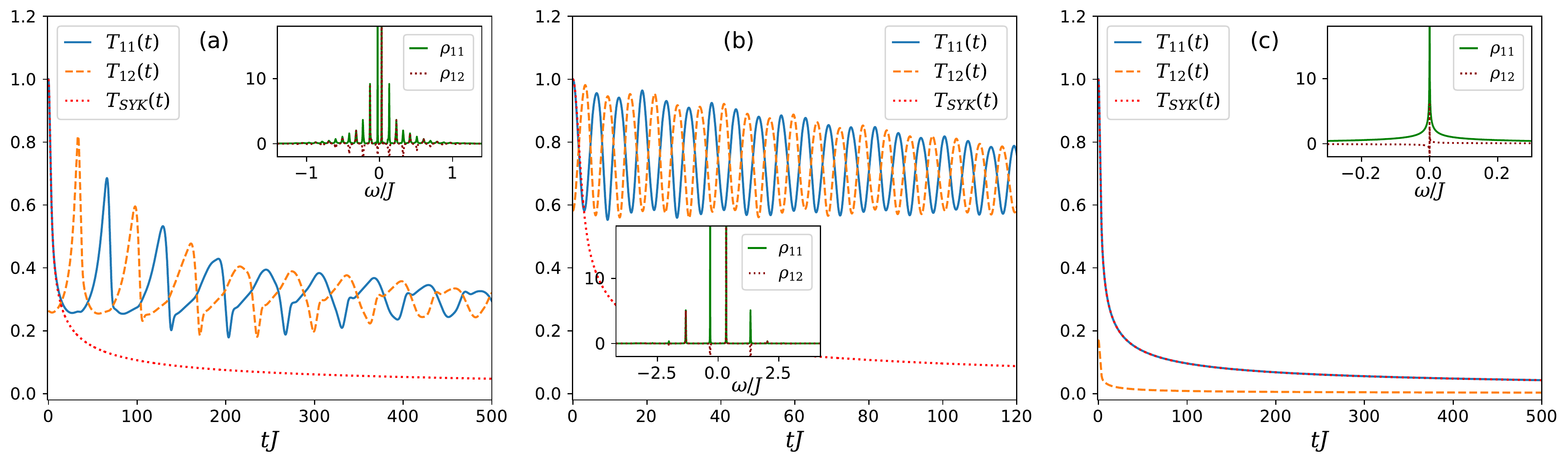}
	\caption{Transmission amplitudes $T_{11,12}(t)$, as in Eq.~\eqref{eq:transmission}, for (a) $\kappa=0.005$ and $\alpha = 0$, (b) $\kappa = 0.005$ and $\alpha=-1$, (c) $\kappa=0.1$ and $\alpha=1$, with $ T/J = 10^{-4}$. All cases show an initial decay of $T_{11}(t)$ following the SYK power-law behavior $T_\mathrm{SYK}(t)$ (with $\kappa=\alpha=0$). For parameters as in (a,b) that lead to a gapped phase (see the spectral functions in the respective insets) we find oscillations in both transmissions $T_{11,12}(t)$ that are out of phase as expected in the wormhole scenario~\cite{Plugge2020}. For parameters as in (c) the transmission $T_{11}(t)$ instead tracks the SYK curve, reflecting a black hole-like decay of two-point functions in system $1$. The spectral function $\rho_{11}(\omega)$ exhibits SYK behavior, while $\rho_{12}\neq 0$ is anti-symmetric about $\omega =0$. Hence $T_{12}(t)$ is non-zero and shows a similar decay, reflecting persistent correlations between the two systems.}
	\label{fig:revivals}
\end{figure*}
When introducing a non-zero chemical potential $\mu$, the system is not necessarily charge neutral. For a single cSYK model this results in a spectral asymmetry or a `twist' ${\cal E}$ in the conformal limit of the imaginary-time Green's functions~\cite{Sachdev2015,Gu2020}, see Eq.~\eqref{eq:correlator_SYK}. The corresponding U(1) charge density $\mathcal{Q} $ can be read off from the value of the imaginary-time Green's functions near $\tau=0$,
\begin{equation}
    G_{11}(0^+) = \frac{1}{2} - \mathcal{Q} \quad , \quad G_{11}(0^-) = -\frac{1}{2} - \mathcal{Q}.
    \label{eq:charge_zerotime} 
\end{equation}

In Fig.~\ref{fig:munonzero_MQ} we show the U(1) charge, obtained numerically from Eq.~\eqref{eq:charge_zerotime} as a function of parameters $\kappa$ and $\mu$.  We find that the gapped wormhole phase is stable to the inclusion of a finite chemical potential $\mu$, even though $\mu$ breaks the microscopic anti-unitary symmetry $P$ used to define the TFD state. The charge density of the wormhole phase remains zero throughout. Increasing $\mu$ drives the system into a gapless phase with a tunable charge density $\mathcal{Q} \in [-\frac{1}{2}, \frac{1}{2}]$ and then finally to a gapped, polarized state with the maximal charge $\mathcal{Q} = \pm 1/2$. The transition to the polarized state is of first order, where the extensive entropy of the non-Fermi liquid phase jumps to zero, similar to the transition seen in Ref.~\cite{Tatsuoetal2018}.

\subsection{Real-time dynamics}\label{sec:realtime}

In order to probe the dynamical behavior of the model we now switch to real-time representation of the saddle-point equations (see Appendix~\ref{app:realtime} for details). Following Ref.~[\onlinecite{Plugge2020}] we focus on the transmission amplitude
\begin{equation}
    T_{ab}(t) = 2 \left| G^>_{ab}(t) \right| ~ , ~ G^>_{ab}(t) =  \frac{\theta(t)}{N} \sum_j \braket{c_{ja}(t) c^\dagger_{jb}(0)}
    \label{eq:transmission}
\end{equation}
which expresses the probability amplitude of recovering a fermion in system $a$ at time $t$ after inserting the corresponding fermion in system $b$ at time $0$, averaged over all fermionic modes $j$ in the system. Fig.~\ref{fig:revivals}a shows the transmission amplitudes for small $\kappa$ in the low temperature regime $T/J = 10^-4 \ll \kappa/J$. They exhibit sharply peaked revival oscillations in both $T_{11}$ and $T_{12}$ that are out-of-phase, consistent with the propagation of fermions back and forth between the two chaotic systems. As with the Majorana case~\cite{Plugge2020} we find that the sharp revivals rely on a tower of equally-spaced states in the spectral function, see the inset in Fig.~\ref{fig:revivals}a, which occur at harmonics of the gap $\sim \kappa^{2/3}$. The overall decay of oscillations is due to the width of those spectral peaks, which increases when going to higher frequencies and/or temperatures~\cite{Zhang2020}. By comparison, for temperatures above the first-order transition shown in Fig.~\ref{gapscalingcSYK}d we observe a smooth, power-law decay $\sim |t|^{-1/2}$ of the transmission amplitude characteristic of the SYK non-Fermi liquid phase.

Based on the results of this Section we conjecture that two identical cSYKs models coupled with a weak tunneling term admit a low-temperature phase which is holographically dual to a traversable wormhole. We rely on the following observations: (i) the presence of a TFD ground state with large $\tilde{\beta}$, (ii) a first-order phase transition separating the (presumed) gapped wormhole phase from a gapless cSYK phase at high temperature and (iii) revival dynamics showing the transmission of excitations between the two chaotic subsystems.

\section{\label{sec:alphamodel} Coupling $c$SYK models with interaction terms}

In this section we couple the two cSYK models with four-fermion interactions that conserve charge on each system. We consider the Hamiltonian introduced in the previous section, modified by an extra term
\begin{align}\label{htotal} 
H = H_{\kappa} + \alpha \sum_{i,j,k,l} J_{ij;kl} \left(c_{i1}^{\dagger}c_{j2}^{\dagger} c_{k1} c_{l2} + c_{i2}^{\dagger}c_{j1}^{\dagger} c_{k2} c_{l1}\right),
\end{align}
where the coupling constants $J_{ijkl}$ are identical to those within each cSYK system. Related models have been studied before in the context of symmetry-broken ground states~\cite{KimKlebanov2019} and superconducting instabilities of SYK non-Fermi liquid phases~\cite{Chowdhury2019}. An additional motivation to study such an interaction term, as explained in more details in Sec.~\ref{sec:graphene}, is that it naturally arises in proposed physical realizations of the SYK model in graphene flakes~\cite{Chen2018}. 
\begin{figure*}
\includegraphics[width=\textwidth]{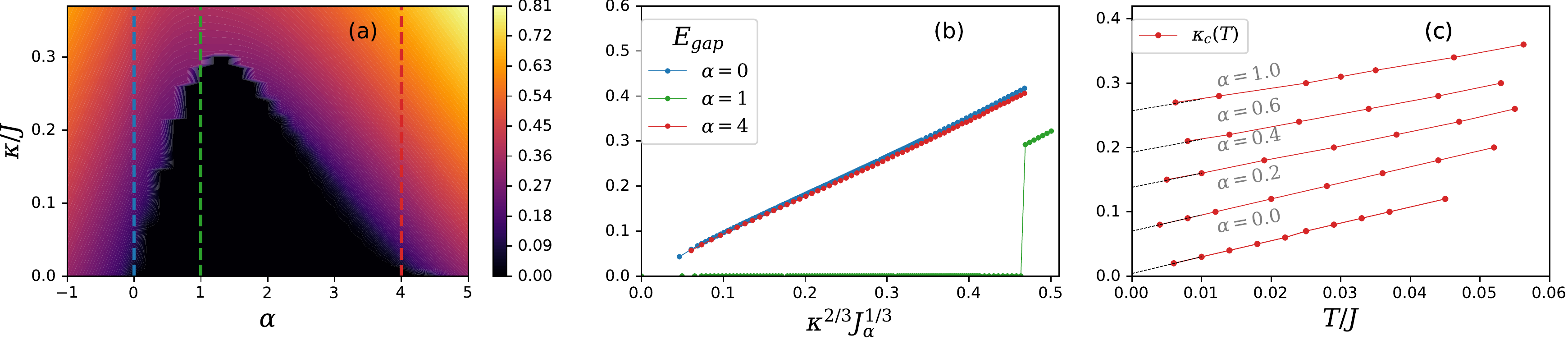}
\caption{\label{gapscalingmu} (a) Spectral gap as a function  of parameters $\kappa$, $\alpha$ in Eq.~(\ref{htotal}), for temperature $T/J =10^{-4}$ and at charge neutrality $\mu=0$. (b) Gap scaling as a function of $\kappa$ for different values of $\alpha$, with $J_\alpha^2 = J^2(1+\frac{1}{2} \alpha^2)$ taken constant across all cases to allow a direct comparison. (c) First-order phase transition lines for $\alpha$ in the range from $0$ to $1$, which separate the gapless and gapped phases of the coupled cSYK models.}
\end{figure*}
The additional $\alpha$ term alters the symmetries of the model, resulting in different degeneracies. For $\alpha \neq 0$ but $\kappa = 0$, $P_1$ and $P_2$ are no longer symmetries but the combined anti-unitary symmetry $P_{12} = P_{1}P_{2}$ remains, where $P_{12}^2 = (-1)^{N(2N-1)}$. This symmetry guarantees a two-fold degeneracy for odd $N$ (see Table~\ref{tab:deg}). 
As before, degeneracies are lifted at $\kappa \neq 0$ as the tunneling term breaks the anti-unitary symmetry for odd $N$, $P_{12}^{-1} K P_{12} = -K$.

The large-$N$ saddle-point equations are obtained in a similar way as Eqs.~(\ref{SDeqn}), with details  delegated to App.~\ref{app:largeN}. The equations for the Green's functions $G_{ab}(i \omega_n)$ are unchanged while the expressions for the self-energies $\Sigma_{ab}(\tau)$ acquire additional terms
\begin{align} 
\label{SDalpha1} 
\Sigma_{11}&(\tau) = - J^2 \Big[ \left(1+\frac{\alpha^{2}}{2}\right) G_{11}^{2}(\tau) G_{11}(-\tau)  \\ &-2 \alpha \ G_{11}(\tau) G_{12}(\tau) G_{12}(-\tau) - \frac{\alpha^2}{2} G_{11}(-\tau) G_{12}^{2}(\tau) \Big], \nonumber \\
\label{SDalpha2} 
\Sigma_{12}&(\tau) = J^2 \Big[ \left(1+\frac{\alpha^{2}}{2}\right) G_{12}^{2}(\tau) G_{12}(-\tau)  \\ &-2 \alpha G_{11}(\tau) G_{11}(-\tau) G_{12}(\tau) - \frac{\alpha^2}{2} G_{12}(-\tau) G_{11}^{2}(\tau) \Big] . \nonumber
\end{align}
The low-temperature phase diagram of this model, obtained from the self-consistent numerical solution of the above equations, is analyzed in Fig.~\ref{gapscalingmu}. At charge neutrality we find two phases whose properties are discussed in the next subsection: a gapless phase with tunable charge density similar to the cSYK non-Fermi liquid, and a gapped phase that is adiabatically connected to the $\alpha=0$ wormhole solution of Sec.~\ref{sec:model}. The gapped phase persists down to $\kappa=0$ for $\alpha < 0$ or $\alpha > 4$ through a U(1) symmetry-breaking mechanism, where a finite expectation value $\braket{K}$ is generated spontaneously. We discuss the physics away from charge neutrality in Sec.~\ref{sec:twoblackholes_finitemu}.

\subsection{Phase diagram at charge neutrality}
\label{sec:phase_diagram}

The low-temperature phase diagram of the model, Eq.~(\ref{htotal}) consists of two phases near charge neutrality $\mu=0$.
For $\alpha < 0$ or $\alpha > 4$ the system is in a gapped phase for all values of $\kappa$, as indicated by the exponential decay of the two-point correlators $G_{ab}(\tau)$ at late times $1 \ll \tau J \ll J/T$. In Fig.~\ref{gapscalingmu}a,b we show the gap extracted from that exponential decay at low temperature $T/J = 10^{-4}$ as a function of $\kappa$ and $\alpha$. For $0 \leq \alpha \leq 4$, we find that a gapless phase survives for a range of $\kappa$ inside a dome-shaped region, where the correlator $G_{11}(\tau) \sim \tau^{-1/2}$ decays as a power-law with the same exponent as in the SYK phase.
For $\alpha = 4$ a gap opens for any $\kappa \neq 0$, similarly to the $\alpha=0$ case analyzed in Sec.~\ref{sec:model}. This can be easily understood from the saddle-point equations (\ref{SDalpha1},\ref{SDalpha2}). Using the symmetry of the two-point correlators at charge neutrality, $G_{11}(-\tau) = -G_{11}(\tau)$ and $G_{12}(-\tau) = G_{12}(\tau)$ we see that at $\alpha=4$ the saddle-point equations reduce to 
\begin{align}\label{eq:sigmas-alpha4}
\Sigma_{11}(\tau) &= - 9 J^2  G_{11}^{2}(\tau) G_{11}(-\tau) \\
\Sigma_{12}(\tau) &= 9 J^2 G_{12}^{2}(\tau) G_{12}(-\tau) \label{eq:sigmas-alpha42}
\end{align}
which are just the equations for two decoupled cSYK models ($\alpha = 0$) but with a renormalized $J_\mathrm{\alpha} = 3J$.
When turning on a finite $\kappa$ we thus expect the same low temperature `wormhole' physics as for $\alpha=0$, but with a renormalized gap $E_{\rm gap} \sim \kappa^{2/3} J_\mathrm{\alpha}^{1/3}$.
We verify this scaling from our numerical simulations, as shown in Fig.~\ref{gapscalingmu}b . The mapping between $\alpha=0$ and $\alpha = 4$ is reminiscent of the duality that exists in the analogous Majorana model in Ref.~[\onlinecite{KimKlebanov2019}]. However, here the duality is only emergent in the large-$N$ saddle-point equations, and is \emph{not} present in the microscopic Hamiltonian.

We then compute the free energy of the model, using the same approach as in Sec.~\ref{sec:model}, Eq.~\eqref{eq:free_energy}. The free energy shows hysteresis across the phase transition between the gapless and gapped phases for $0\leq \alpha \leq 4$. The phase transition lines for various $\alpha$ are shown in Fig.~\ref{gapscalingmu}c. As $\alpha$ increases the first-order transition lines move up the $\kappa$ axis and have a non-zero intercept, such that the gapless phase extends down to zero temperature. Thus for $0 \leq \alpha \leq 4$, the first-order Hawking-Page phase transition occurs at zero temperature upon varying the tunneling strength. This is also indicated by the discontinuous jump in the gap magnitude across the transition shown in Fig.~\ref{gapscalingmu}b.

\begin{figure}
\begin{center}
\includegraphics[width=0.95\columnwidth]{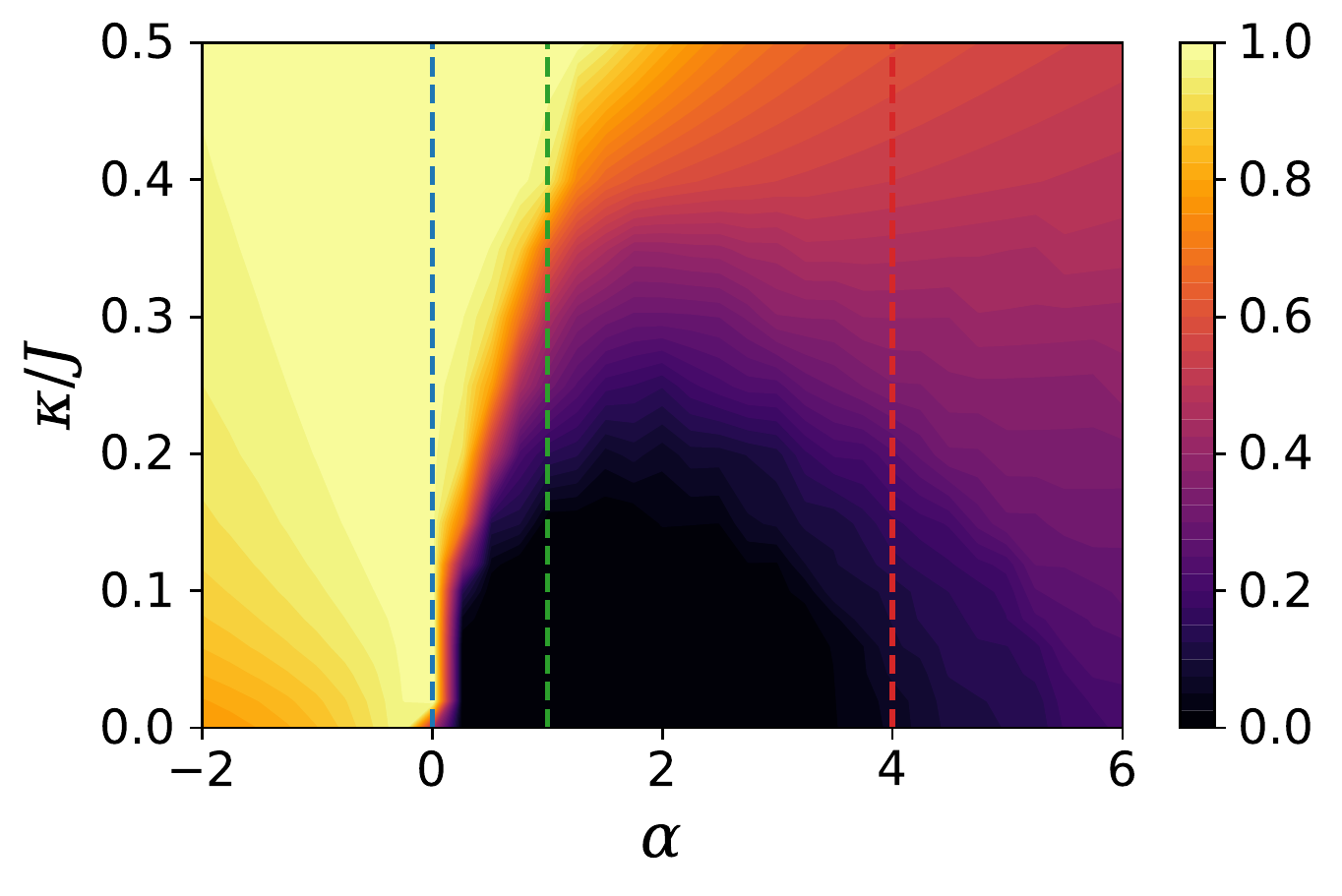}
\end{center}
\caption{Overlap between the ground state $\ket{\Psi_0}$ of the coupled cSYK models, Eq.~(\ref{htotal}), at charge neutrality and the best-fit $\rm{TFD}$ state. Results are obtained from exact diagonalization
with $2N = 12$ and averaged over 20 disorder realizations.
\label{fig:ED_alpha} }
\end{figure}

Using exact diagonalization we obtain the overlap between the ground state of the coupled model and the TFDs defined in Eq.~(\ref{TFD}), as shown in Fig.~\ref{fig:ED_alpha}. In the gapped phase the overlap with the TFD decreases continuously when moving away from the $\alpha=0$ `wormhole' line, and sharply drops to zero upon entering the gapless phase. Interestingly, at $\alpha=4$ the ground state is not well approximated by a TFD, a further indication of the duality between $\alpha=0$ and $4$ being only valid in the large-$N$ limit.

\subsection{Spontaneous symmetry breaking}

\begin{figure}
\begin{center}
\includegraphics[width=0.95\columnwidth]{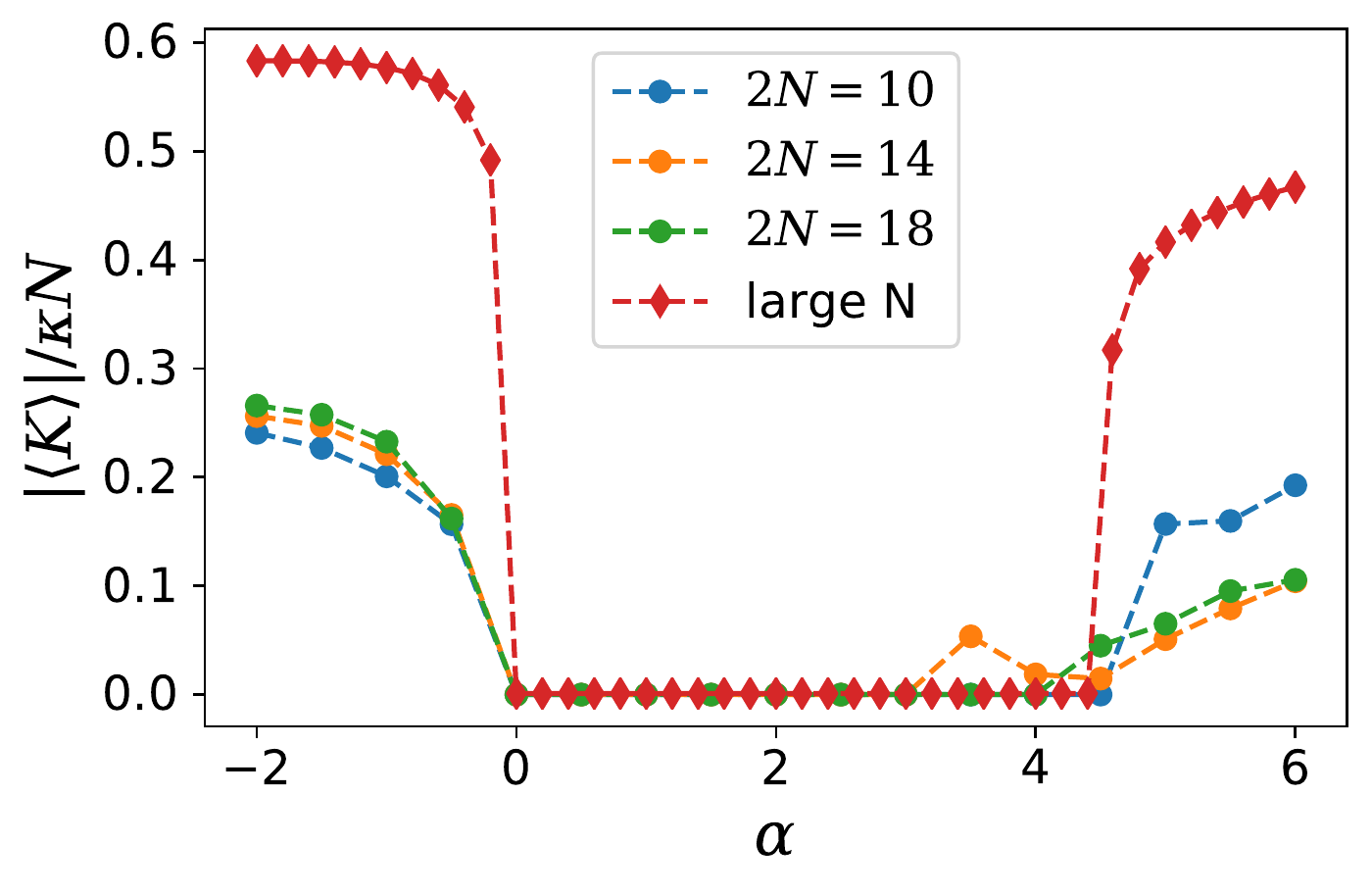}
\end{center}
\caption{Symmetry-breaking mechanism for $\kappa=0$, whereby a finite expectation value of the tunneling operator $\braket{K}/\kappa N$ is generated spontaneously. The large-$N$ result is plotted alongside exact diagonalization results for odd $N$, which represent the eigenvalues of $\braket{K}/\kappa N$ in the ground state manifold. The ED results are averaged over 8 disorder realizations.}
\label{fig:sym_breaking} 
\end{figure}

When $\kappa = 0$ and $\alpha<0$ or $\alpha>4$, the system spontaneously develops a non-zero expectation value for the tunneling operator $K$, which can be read off numerically from $\frac{\braket{K}}{\kappa N} = 2i G_{12}(\tau = 0)$, as shown in Fig.~\ref{fig:sym_breaking}. We therefore conclude that the gap opens at $\kappa=0$ through a spontaneous symmetry-breaking mechanism, from the U(1)$\otimes$U(1) charge conjugation symmetry down to U(1). This spontaneous U(1) symmetry breaking is accompanied by a gapless Goldstone mode reflecting phase fluctuations of the corresponding order parameter~\cite{Klebanov2020}.

For finite $N$ the spontaneous symmetry breaking can be analyzed using exact diagonalization, providing  a simple explanation of the phase transition observed at $\kappa=0$. For even $N$ the ground state is unique and we always find that $\braket{K}=0$. For odd $N$ however, we have two degenerate ground states which, in the gapped phase, are located in the $q=0$ sector (see Table~\ref{tab:deg}). We can perform a basis rotation in this twofold degenerate space to obtain two eigenvectors of $K$ with opposite eigenvalues, as shown in Fig.~\ref{fig:sym_breaking}. The system can thus spontaneously choose a ground state which breaks U(1)$\otimes$U(1) symmetry, as in the saddle-point result. In the process the anti-unitary symmetry is also spontaneously broken as $P_{12}^{-1} K P_{12} = -K$ for odd $N$. In contrast, in the gapless phase (for $0 \leq \alpha \leq 4$) the two ground states are in charge sectors $q = \pm 1$ and have $\braket{K}=0$ since $K$ conserves charge. There is thus no possible symmetry breaking, in accordance with the large $N$ result. 

\subsection{Revival dynamics}
\label{sec:revivals-alpha}

The transmission amplitudes [Eq.~\eqref{eq:transmission}] for non-zero $\alpha$ at low temperature $T/J = 10^{-4}$ are shown in Fig.~\ref{fig:revivals}b,c. For $\alpha = -1$ and small $\kappa$, deep inside the gapped phase, we again find revival oscillations. Those are notably less sharp than at the MQ point $\alpha=0$, consistent with the observation that the ground state is not well approximated by a TFD. 
The reason is that the gap remains large as $\kappa \rightarrow 0$. Therefore, there is only a small number of states in the conformal tower (at harmonics of the gap) that can fit within the energy scale $J$ which limits the conformal scaling behavior~\cite{Plugge2020}. To this end compare the spectral functions at $\alpha=0$ (Fig.~\ref{fig:revivals}a inset), showing a large number of evenly spaced peaks, and $\alpha=-1$
(Fig.~\ref{fig:revivals}b inset), showing only few and far-spaced spectral peaks. Thus at $\alpha=-1$ the revivals are controlled by a few spectral peaks rather than an extensive tower of states.
For $\alpha=1$ and small $\kappa$ (deep inside the gapless dome) we find a power-law decay of the transmission down to the lowest accessible temperatures, closely tracking SYK behavior, but now with $T_{12}(t)\neq 0$ also decaying as a power-law. The spectral function $\rho_{11}$ (Fig.~\ref{fig:revivals}c inset) shows gapless power-law behavior as expected in the SYK phase, however $\rho_{12} \neq 0$ indicates the presence of correlations between the two subsystems.

\subsection{Conformal solution and SU(2) symmetry}
\label{sec:gapless}

\begin{figure}
\centering
\includegraphics[width=0.95\columnwidth]{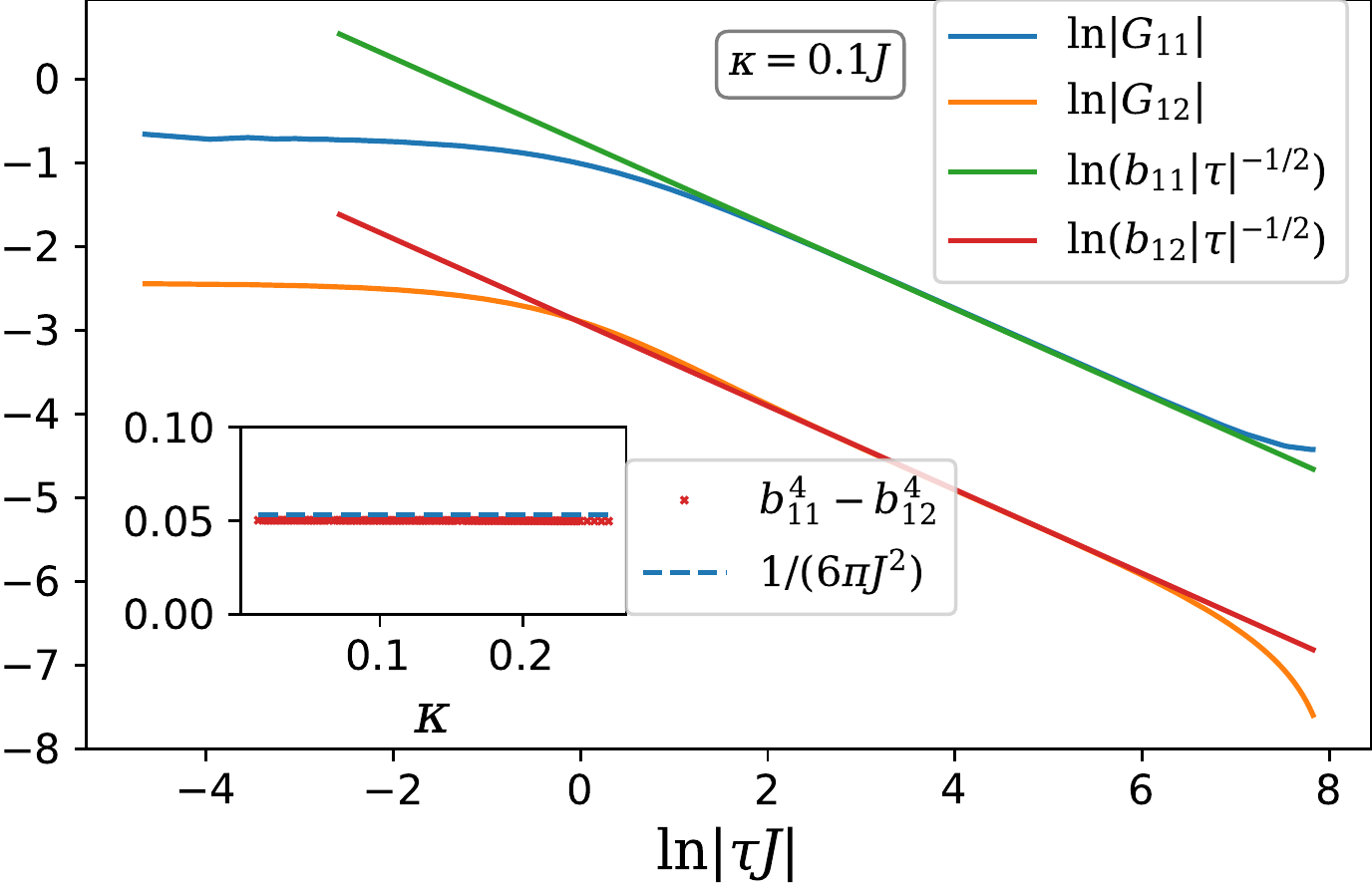}
\caption{Imaginary-time correlators $G_{11}(\tau)$ and $G_{12}(\tau)$ versus time, on log scales, for $\alpha = 1,~\kappa=0.1J$ and low temperature $T/J = 0.002$. Fitting to the conformal scaling form for $G_{ab}(\tau)$ gives the coefficients $b_{ab}$. The inset shows the difference of fourth powers of the coefficients as function of $\kappa$, verifying the analytical solution discussed in the text.
}
\label{fig:correlator_alpha}
\end{figure}

We now consider whether a low-energy conformal solution of the saddle-point equations (\ref{SDalpha1}), (\ref{SDalpha2}) can be found to describe the gapless phase. When $\kappa =0$, we have $G_{12}(\tau)=0$ and the saddle-point equations can be solved at low energies to show that $G_{11}(\tau)$ is a conformal cSYK correlator with $J^{2}$ replaced by $J_{\alpha}^2=J^{2}(1+\frac{1}{2}\alpha^{2})$. Note that this connects smoothly with the behavior at $\alpha = 0,~4$ discussed below Eqs.~\eqref{eq:sigmas-alpha4}-\eqref{eq:sigmas-alpha42}. With $\kappa > 0$, $G_{12}(\tau)$ does not vanish which renders the analytical solution of the saddle-point equations more difficult. In the low energy limit, we find that a conformal solution for both $G_{11}$ and $G_{12}$ is in general not possible, except at the special point $\alpha=1$ discussed in Appendix~\ref{appendix:Conformalsol}. This limit admits a power-law solution with the same power for both correlators, $G_{ab}(\tau) \sim b_{ab} |\tau|^{-1/2}$ in the long-time limit $1 \ll \tau J \ll J/T$, and where the two coefficients are related by $b_{12}^4 = b_{11}^4 - 1/\left( 6 \pi J^2 \right)$. This is demonstrated numerically in Fig.~\ref{fig:correlator_alpha}.
Note that the saddle-point equations alone are not sufficient to fix the coefficients of the two power laws. Instead one must impose a constraint linking microscopic and conformal physics, similar to how U(1) charge enters the conformal solution in the cSYK model~\cite{Georges2001, Sachdev2015} (see Appendix~\ref{appendix:Conformalsol}).

The $\alpha=1$ point is special because it has SU(2) symmetric interactions~\footnote{The full symmetry group at $\alpha=1$ is $ {\rm U(2) = SU(2) \otimes U(1)}$, as the U(1) charge conservation is also present.}. An important consequence is that the tunneling term $K$ is now a symmetry of the model, $[H_{\alpha=1}, K] = 0$. Hence $\braket{K}$ is a conserved quantity that can be tuned by the tunneling parameter $\kappa$, in analogy with the U(1) charge density $\mathcal{Q}$ tuned by the chemical potential $\mu$. However, the two symmetries have different signatures: introducing $\kappa$ generates a non-zero value of $iG_{12}$, but not a twist parameter leading to a spectral asymmetry, as occurs with non-zero $\mu$. Another consequence of the SU(2) symmetry is that the non-interacting ground state $\ket{\rm{TFD}_0}$ of $K$ is an eigenstate of the full model for any $\kappa$. In fact, exact diagonalization shows that for $\kappa > \kappa_c \simeq 0.27 J$ the model admits the $\ket{\rm{TFD}_0}$ state as an \emph{exact} ground state. Because the interaction and tunneling terms commute, the only way to change the ground state is through an energy level crossing, which occurs at the first-order transition at $\kappa_c$ where the gapless phase becomes favored.

\subsection{Moving away from charge neutrality: a tale of two black holes}
\label{sec:twoblackholes_finitemu}

We now discuss the physics away from charge neutrality. 
%
%
\begin{figure}
\includegraphics[width=0.95\columnwidth]{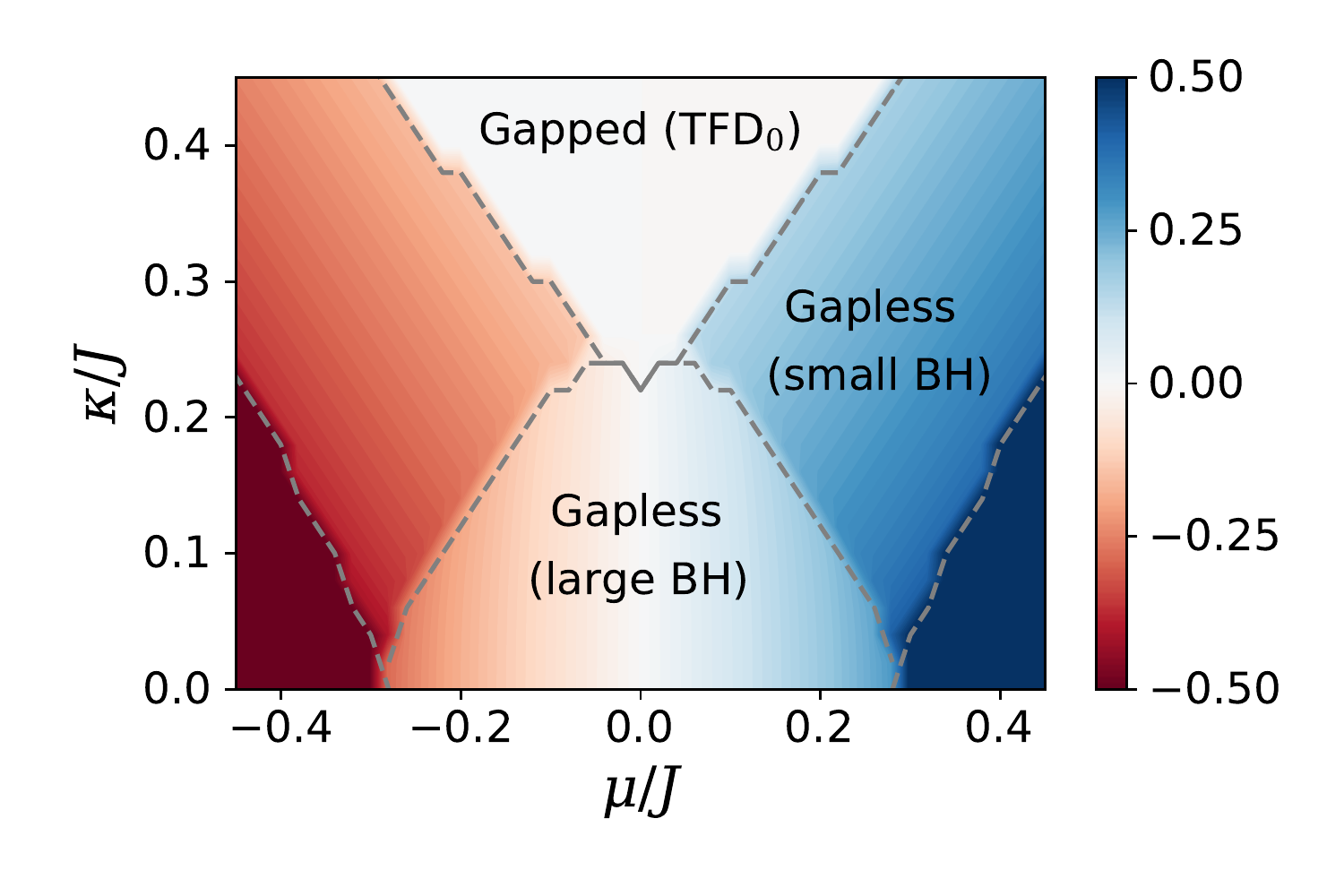}
\includegraphics[width=0.95\columnwidth]{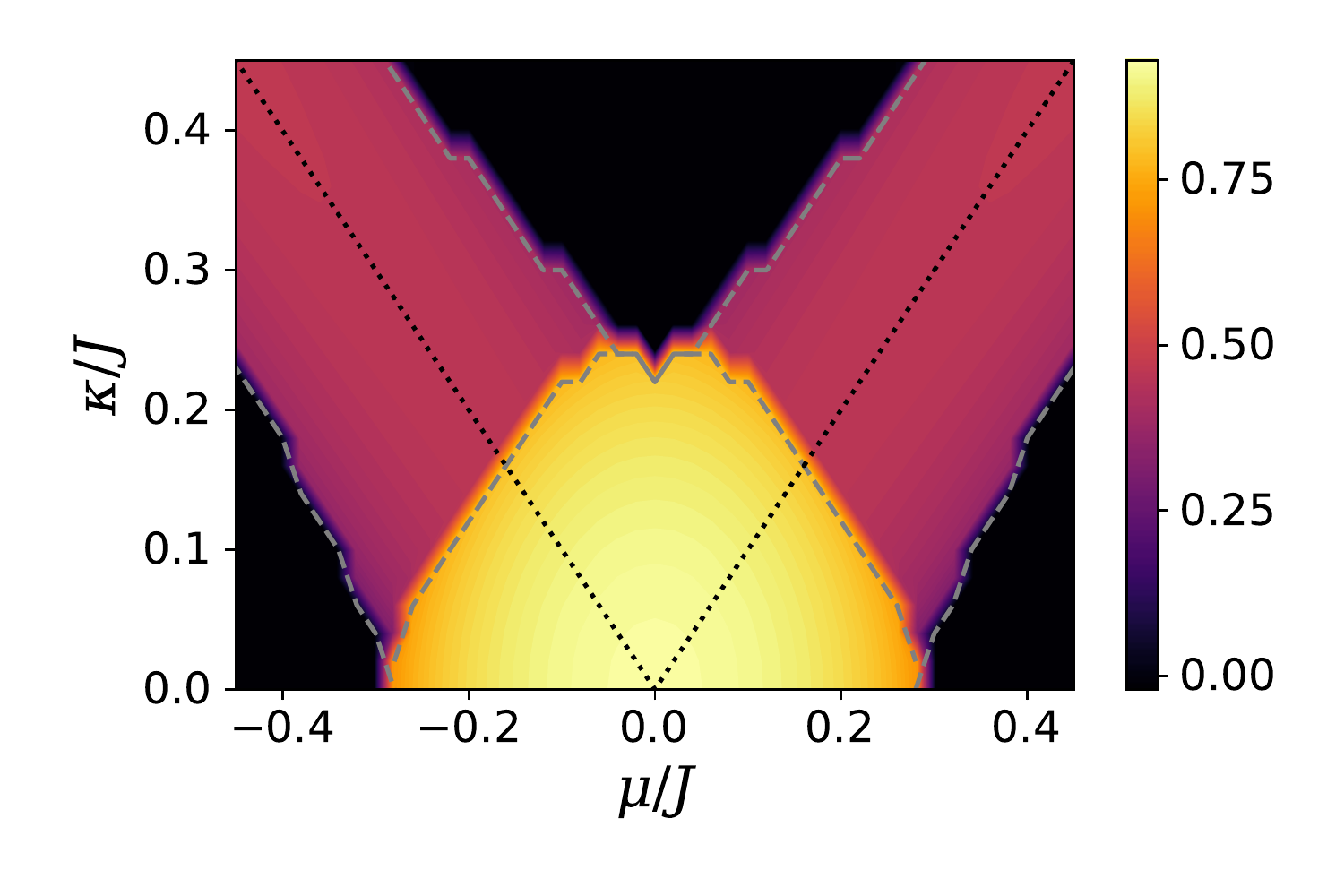}
    \caption{Charge density $\mathcal{Q}$ (top) and residual entropy density $\mathcal{S}_0$ (bottom) as a function of $\kappa$ and $\mu$, at the SU(2) invariant point $\alpha=1$ and low temperature $T/J = 0.002$. The charge density is tunable in the compressible gapless phases. Large $|\mu|$ leads to a gap opening when $\mathcal{Q} = \pm 1/2$, corresponding to a fully polarized state, while large $\kappa$ leads to a charge-neutral gapped phase with a $\ket{\rm{TFD}_0}$ ground state. The two gapless phases show extensive residual entropy consistent with a `large' and `small' black hole with $2N$ and $N$ degrees of freedom, respectively. All phases are separated by first-order phase transitions indicated by dashed lines. The cuts $\mu = \pm \kappa$ discussed in the text are shown by black dotted lines.}
    \label{fig:munonzero_SU2}
\end{figure}
We focus on the SU(2) symmetric point $\alpha=1$ and investigate its low-temperature phase diagram in the $\kappa -- \mu$ plane. To this end we show in Fig.~\ref{fig:munonzero_SU2} the U(1) charge obtained numerically from Eq.~\eqref{eq:charge_zerotime} and the residual entropy density $\mathcal{S}_0$, as a function of parameters $\kappa$ and $\mu$ at low temperature $T/J =0.002$. We recover the known gapped phases discussed above: for large $\kappa$, which admits the non-interacting charge-neutral $\ket{\rm{TFD}_0}$ ground state, and for large $|\mu|$, which corresponds to the polarized $\mathcal{Q} = \pm 1/2$ states. The boundaries of the two gapped phases host first-order phase transitions to gapless non-Fermi liquids, as indicated by the discontinuous jump in entropy density at the phase boundaries. Surprisingly, we find not one but \emph{two} such gapless phases. 
Near charge neutrality, we obtain a phase smoothly connected to the conformal solution discussed above. In this phase both correlators $G_{11}$ and $G_{12}$ show power-law decay which indicates strong correlations between the two subsystems. This phase can be thought of as a single cSYK phase with $2N$ fermions, dual to a `large' black hole comprising all degrees of freedom in the combined system. 

\begin{figure}
\centering
\includegraphics[width=0.95\columnwidth]{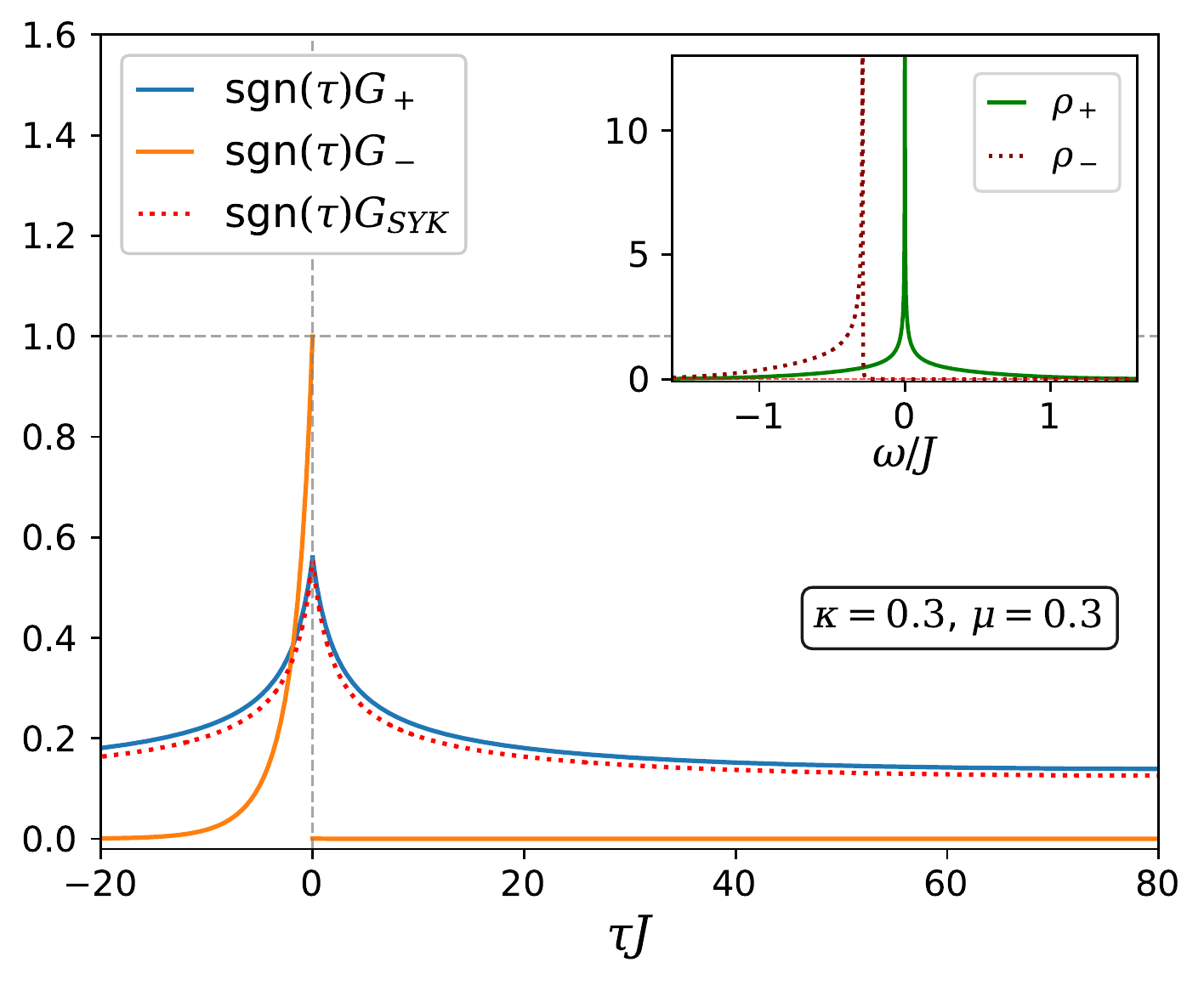}
\caption{Imaginary-time correlators $G_\pm(\tau)$ of fermions $c_\pm$, cf. Eq.~\eqref{eq:Hpm}, for $\alpha = 1,~\kappa=\mu=0.3J$ and $T/J = 10^{-4}$. Inset: corresponding spectral functions $\rho_\pm$. The behavior of $G_+$ and $\rho_+$ closely follows that of a charge-neutral cSYK model, while  $\rho_-$ is gapped with finite spectral weight only at negative frequencies. Hence $G_-(\tau)$ is zero for $\tau>0$ and shows exponential decay for  $\tau<0$.
}
\label{fig:Gpm-kmu}
\end{figure}

Farther from charge neutrality we find another first-order phase transition to a different gapless phase with charge density $\mathcal{Q} \simeq 0.25$ and about half of the residual entropy at charge neutrality. To understand this, note that at $\alpha=1$ where the interactions are SU(2) invariant, we can rotate to a new basis 
$c_{j \pm} = \frac{1}{\sqrt2}(c_{j1} \pm i c_{j2})$, such that
%
%
%
\begin{equation}\label{eq:Hpm}
    H = \sum_{a,b=\pm} \sum_{i,j,k,l} J_{ij;kl} c_{ia}^{\dagger}c_{jb}^{\dagger} c_{ka} c_{lb} - \sum_{a=\pm, j} \mu_a c_{ja}^\dagger c_{ja}.
\end{equation}
In this basis one can interpret the system as two cSYK models with different chemical potentials $\mu_\pm = \mu \mp \kappa$ and SU(2) invariant interactions between them. Let us first focus on the $\kappa = \mu$ line where the chemical potentials are simply $0$ and $2\mu$. When $\mu$ increases, eventually one of the cSYK models undergoes a first-order phase transition to a gapped, polarized state with $\mathcal{Q}=1/2$. At low energies (below the gap), its degrees of freedom thus decouple and we are left with the other cSYK model at charge neutrality $\mathcal{Q} =0$. The combined system thus has exactly $\mathcal{Q} = 1/4$ and $\mathcal{S}_0 = \mathcal{S}_{\rm cSYK}$, as observed in the saddle-point solutions. In Fig.~\ref{fig:Gpm-kmu} we show the spectral functions $\rho_\pm$ for the rotated basis fermions $c_\pm$. We observe a power-law scaling at low frequency in the $\rho_+$ channel while the $\rho_-$ channel is gapped, confirming the argument above. In imaginary time, the corresponding correlators $G_\pm(\tau) = \frac{1}{N} \sum_j \langle \mathcal{T} c_{j\pm} (\tau) c_{j \pm}^\dagger(0) \rangle = G_{11}(\tau) \mp i G_{12}(\tau) $ show power-law and exponential decay, respectively, at long times.

In the vicinity of the $\mu = \pm \kappa$ lines the residual entropy and charge density change smoothly, see Fig.~\ref{fig:munonzero_SU2}, as one half of the system is in a compressible cSYK state while the other half remains gapped. We interpret this phase as dual to a `small' black hole, comprising half of the degrees of freedom of the combined system, with the other half decoupled and frozen into a fully polarized state.  Interestingly, we find that this ``small black hole" phase persists away from $\alpha=1$ even though the SU(2) symmetry allowing the basis change argument is absent.

\section{\label{sec:graphene} Physical realization in graphene flakes}

We now turn to potential physical realizations of the model introduced in this work. As originally described in Ref.~[\onlinecite{Chen2018}], a promising platform for realizing cSYK physics is a mesoscopic graphene flake under a perpendicular magnetic field $\mathbf{B}$, with the chemical potential $\mu$ lying within the zeroth-Landau level (LL$_0$). An Aharonov-Casher argument~\cite{Aharonov} implies that LL$_0$ remains sharp as long as the chiral (sublattice) symmetry of the model is unbroken, thus forbidding two-fermion terms that would destroy the non-Fermi liquid physics at low energies. Disorder that preserves this chiral symmetry (such as an irregular boundary) then imprints disorder on the LL$_0$ wavefunctions \emph{without lifting their degeneracy}, leading to random and all-to-all interactions between them.

Due to the negligible spin-orbit coupling in clean graphene, it is reasonable to assume identical wavefunctions for the two spin components.  This should still hold in the presence of non-magnetic disorder (such as an irregular boundary) which preserves the SU(2) symmetry of Coulomb interactions. The graphene flake setup thus naturally leads to two \emph{identical} copies of the cSYK model, one for each spin component. In Ref.~[\onlinecite{Chen2018}] the authors argued that the Zeeman splitting obtained by applying a magnetic field to the sample generates a large spin gap (augmented by exchange interactions), which effectively reduces the problem to a single cSYK model.

In this Section we revisit this analysis by looking more carefully at the role of spin in the above proposal. Using a mapping to the model studied in Sec.~\ref{sec:alphamodel} we conclude that the graphene flake model with a \emph{weak} Zeeman splitting is in a gapless cSYK phase with fermion scaling dimension $1/4$ and tunable charge density. In contrast to expectations that a strong Zeeman splitting should give rise to a cSYK phase~\cite{Chen2018}, we find that it instead leads to a gapped phase with an exact $\ket{\text{TFD}_0}$ ground state.

The Coulomb interactions between electrons in the graphene flake read
\begin{align} 
H_{\rm int} = \frac{1}{2} \sum_{\mathbf{r},\mathbf{r}'}\rho_{\mathbf{r}} V(\mathbf{r}-\mathbf{r}') \rho_{\mathbf{r}'} 
\label{eq:Coulomb}
\end{align}
where $\rho_{\mathbf{r}} = \rho_{\mathbf{r}\uparrow} + \rho_{\mathbf{r}\downarrow}$ is the total charge density at a point $\mathbf{r}$ in space and $V(\mathbf{r}-\mathbf{r'})$ is the screened Coulomb potential. The electronic charge densities can be expressed in terms of the eigenfunctions $\phi_j({\bm r})$ of the non-interacting Hamiltonian which, neglecting spin-orbit coupling effects, are independent of spin $\sigma = \uparrow, \downarrow$,
\begin{align}
    \rho_{\mathbf{r} \sigma} = c^\dagger_{\mathbf{r} \sigma} c_{\mathbf{r} \sigma} = \sum_{ik} \phi_{i}^*(\mathbf{r}) \phi_{k}(\mathbf{r}) c_{i \sigma}^\dagger c_{k \sigma}
    \label{eq:charge_density}
\end{align}
Projecting to the LL$_0$ wavefunctions, Eqs.~(\ref{eq:Coulomb},
\ref{eq:charge_density}) lead to a (normal-ordered) interaction Hamiltonian with all-to-all couplings $H_{\rm int} = \sum_{ijkl} J_{ij;kl} c_{i \sigma}^\dagger c_{j \sigma'}^\dagger c_{k \sigma} c_{l \sigma'}$ and spin-independent coupling constants
\begin{align} 
 J_{ij;kl} = - \frac{1}{2} \sum_{\mathbf{r}, \mathbf{r'}} \phi_i^*(\mathbf{r}) \phi_k(\mathbf{r})  V(\mathbf{r} - \mathbf{r'}) \phi_j^*(\mathbf{r'}) \phi_l(\mathbf{r'}).
\end{align}
Assuming spatially random wavefunctions and strong screening these become complex random Gaussian variables~\cite{Chen2018} (see Ref.~[\onlinecite{LH2018}] for a discussion of varying screening lengths in a related model).

Adding the Zeeman term, the Hamiltonian describing the low-energy physics of the graphene flake becomes
\begin{align} 
H =  \sum_{ijkl} \sum_{\sigma, \sigma'} J_{ij;kl} c_{i \sigma}^{\dagger}c_{j \sigma'}^{\dagger}c_{k\sigma}c_{l\sigma'} + g \mu_{B}B \sum_{i}(n_{i\uparrow}- n_{i\downarrow}),
\label{eq:graphene}
\end{align}
where $g \sim 2$ and $\mu_B$ are the Land\'e factor and the Bohr magneton, respectively. The Coulomb interactions in this model are invariant under SU(2) rotations in spin space. Following the analysis in Sec.~\ref{sec:twoblackholes_finitemu} we can perform a basis change $c_{j,\uparrow \downarrow} = \frac{1}{\sqrt{2}} ( c_{j1} \pm i c_{j2} )$, which corresponds (up to a gauge transformation) to a rotation by $\frac{\pi}{2}$ along the $x$-axis in spin space,
\begin{align}
H =& \sum_{ijkl} \sum_{a,b} J_{ij;kl} c_{i a}^{\dagger}c_{j b}^{\dagger}c_{k a}c_{l b} 
+ ig\mu_{B}B \sum_{i} (c_{i 1}^{\dagger} c_{i 2} - c_{i 2}^{\dagger} c_{i 1}).
\label{eq:graphene_rotated}
\end{align}
This has the same form as the model in Eq.~(\ref{htotal}) with $\alpha=1$, if we identify $g\mu_{B} B$ with the tunneling amplitude $\kappa$. As mentioned in the previous section, the SU(2) symmetric interactions commute with the tunneling term. In the original basis of Eq.~\eqref{eq:graphene}, this is easily seen by noting that the Zeeman term is proportional to the total spin projection $S^{\rm tot}_z = \sum_i \sigma_i^z$.

Using this mapping we thus expect that the graphene flake remains in the gapless cSYK non-Fermi liquid phase up to a threshold value $\kappa_c = g\mu_B B_c \sim 0.27 J$. Above the critical field strength $B_c$ the system becomes gapped and the ground state is the infinite-temperature TFD state $|I\rangle$, which in the graphene basis is a fully spin-polarized state, $\prod_j c^\dagger_{j\downarrow} \ket{0}$.  In Ref.~[\onlinecite{Chen2018}] the authors estimate that for $B \sim 20$ T the Zeeman splitting $g\mu_B B \sim 2.4$ meV while the Coulomb interaction strength $J \sim 25$ meV, thus placing the system within the gapless phase. Given various uncertainties in these estimates $J$ could well be smaller in which case the system would realize the gapped phase indicated in Fig.\ \ref{fig:munonzero_SU2}. Further, the power-law scaling characteristic of the conformal regime is expected for temperatures $T \ll J_{\alpha} \sim \frac{3}{2} J$ at the SU(2) symmetric point $\alpha = 1$, corresponding to $T \ll 430$ K which should enable exploration of the low temperature regime.

As shown in Sec.~\ref{sec:twoblackholes_finitemu} the cSYK non-Fermi liquid phase for $B=0$ is stable against inclusion of a chemical potential up to a threshold value $\mu_c$ which corresponds to adding charge density $\mathcal{Q}=1/2$, filling all the states in LL$_0$. For $B\neq0$ the system first transitions to an intermediate non-Fermi liquid phase corresponding to a small black hole. This transition could be explored by tuning the chemical potential in the graphene flake by external gates. Experimentally, the non-Fermi liquids could be distinguished from each other, and from the gapped phases at large $\kappa$ or large $\mu$, by measuring their charge compressibility $\partial Q/\partial \mu$~\cite{Davison2017,Gu2020} or their spectral function through spin-polarized scanning probe techniques. Indeed, the small black hole phase corresponds to an SYK-type non-Fermi liquid for one spin component and a gapped, polarized state for the other. This should be contrasted with the large black hole phase which consists of a non-Fermi liquid in both spin components, 
and the fully gapped states which comprise either filled or empty spin-polarized Landau levels
.

An important caveat of our analysis is that in Eq.~\eqref{eq:graphene_rotated} the coupling constants $J_{ij;kl}$ are only restricted to be antisymmetric under exchanging $(i \sigma , j \sigma)$ and $(k \sigma , l\sigma)$ -- with the same spin component -- by fermionic commutation relations. This is a weaker requirement than the full antisymmetry present for the Majorana SYK model and assumed in this work (the same assumption was made in Ref.~[\onlinecite{Chen2018}]). For example, interaction terms with two pairs of matching indices, corresponding to direct (density) interactions or to exchange (spin) interactions \cite{altlandsimons2010}
are excluded from our model. A detailed analysis of such effects is left for future work.

\section{Conclusion and Outlook}

In this work we generalized the `eternal traversable wormhole' construction of Maldacena and Qi~\cite{maldacenaqi2018} to a system of coupled complex SYK models with a global U(1) charge symmetry. We explained how to define the TFD state in the presence of a U(1) symmetry, and showed that the model admits a gapped phase with a charge-neutral ground state close to a TFD. Whether the weak-tunneling and low temperature limit of the model admits a gravitational dual similar to the wormhole of Ref.~[\onlinecite{maldacenaqi2018}] remains an intriguing open question for the high-energy community. The presence of a gapped ground state close to a large $\tilde{\beta}$ thermofield double, a first-order Hawking-Page phase transition to a gapless cSYK non-Fermi liquid at high temperature and sharp revivals in fermion transmissions are however highly suggestive. 

Further, we considered the effect of four-fermion interactions between the two cSYK models that are disordered identically to the interactions within each system. We explored the phase diagram of the system as a function of tunneling, interactions and chemical potential in Figs.~\ref{gapscalingmu}, and \ref{fig:munonzero_SU2}. At low temperature we obtain three non-trivial phases: a gapped, charge-neutral phase which is adiabatically connected to the conjectured wormhole and two gapless, compressible non-Fermi liquid phases which describe either a `large' or a `small' (i.e. with half of its degrees of freedom gapped out) charged black hole with an AdS$_2$ horizon. All phases are separated by first-order phase transitions exhibiting extensive residual entropy jumps. The transition out of the gapped wormhole phase can be understood as a Hawking-Page transition, as it separates a black hole from a phase appearing locally thermal, a consequence of its TFD ground state. The phase diagram also contains the special case $\alpha=1$ with SU(2)-symmetric interactions which admits a conformally invariant solution at low energies, and is directly relevant to the graphene flake proposal of Ref.~[\onlinecite{Chen2018}].

We conclude by highlighting a few caveats of our analysis and point out interesting directions for further work. First, in order to define the anti-unitary particle-hole symmetry $P$ which enables the TFD state construction, we restricted the model to only contain interactions that are completely antisymmetric in the indices $i,j,k,l$. In the large-$N$ limit these should dominate as their number scales as $N^4$ (in contrast, the number of terms with one or two pairs of identical indices, of the form $J_{ij;ik}$ or $J_{ij;ij}$, respectively scales as $N^3$ and $N^2$). However, for mesoscopic realizations of SYK physics with finite $N$ these terms could be important; understanding their effect will be an important step towards connecting our results with ongoing experimental efforts.

A different approach to define a TFD for cSYK models could be to rely on an anti-unitary \emph{time-reversal} (rather than particle-hole) symmetry. Time-reversal symmetry is obviously broken for a single cSYK model (as manifest by the coupling constants $J_{ijkl}$ being complex), but can be restored globally by considering a pair of time-reversed cSYK models. We anticipate that the TFD construction, saddle-point physics and physical realizations will be different in this case, and leave its detailed study for future work. Another interesting topic concerns the quantum chaotic properties of our model: how does scrambling, as captured through out-of-time-ordered correlators, behave across the Hawking-Page transition which separates the wormhole and black hole phases?

Finally, in light of the rich phase diagram of our model, it is interesting to ask about potential physical realizations for generic $\alpha \neq 1$. In the graphene flake proposal, $\alpha=1$ is enforced by the SU(2) symmetry of Coulomb interactions. However, in other model systems where the two subsystems are realized by two surfaces, such as multilayer graphene~\cite{LH2019} or a topological insulator flake, one could imagine changing the distance between the surfaces as a way to tune the ratio of inter-system to intra-system interactions $\alpha$. In this way one could potentially explore the Hawking-Page phase transition between the wormhole and black hole phases discussed in this work.

\emph{Note added. --} Recently, an independent study of the spontaneous U(1) symmetry breaking in coupled complex SYK models, focusing on the $\alpha$ term, was posted~\cite{Klebanov2020}. Our results match where they overlap.

\section*{Acknowledgments}

We are grateful to Oguzhan Can, Chengshu Li, Moshe Rozali and Xiao-Liang Qi for illuminating discussions. This research was supported in part by NSERC, CIfAR, the Heising-Simons Foundation, the Simons Foundation, and National Science Foundation Grant No. NSF PHY-1748958.

\bibliography{ref}

\onecolumngrid
\appendix

\section{\label{appendix:TFD}TFD construction}
Consider the occupation number basis in the Fock space of $2N$ complex fermions in the doubled system. The product state 
\begin{equation}
    \ket{I} = \prod_{i=1,N} \frac{1}{\sqrt{2}}(|1\rangle_{1}|0\rangle_{2} - e^{-i\phi}|0\rangle_{1}|1\rangle_{2})_{i} 
\end{equation} 
is a unique ground state of the tunneling Hamiltonian term, $K= \kappa \sum_{i} c_{i1}^{\dagger}c_{i2} + \mbox{H.c.}$ with $\kappa$ a complex coefficient $|\kappa| e^{i\phi}$. Showing that the overlap $\braket{I | \rm TFD_0} = 1$, where $\ket{ \rm TFD_0}$ is defined in Eq.~\eqref{iTFD}, is equivalent to showing that the expectation value of $K$ in the TFD state has minimum value i.e. $-|\kappa| N$. Consider the general definition of TFD in Eq.~\eqref{e1} at $\tilde\beta=0$ and evaluate the expectation value
\begin{align}\label{eq:TFDconvention1} 
\braket{K} = \braket{ {\rm TFD}_0 | K | {\rm TFD}_0 } &= \frac{1}{2^N}\sum\limits_{q,q'}\sum\limits_{m,n}\langle \bar{n}_{-q'}|_{2} \otimes \langle n_{q'}|_{1} \left(\kappa \sum_{i} c_{i1}^{\dagger}c_{i2} + \mbox{H.c.} \right)|m_{q}\rangle_{1}\otimes |\bar{m}_{-q}\rangle_{2}.
\end{align}
We insert an identity operator using complete set of basis states with appropriate normalization,
$$I = \sum_{q'', q'''} \sum_{m', n'} \left(|m'_{q''}\rangle_{1} \otimes |n'_{q'''}\rangle_{2}\right) \left(\langle n'_{q'''}|_{2} \otimes \langle m'_{q''}|_{1}\right)$$
to separate the operators acting on system $1$ and $2$,
\begin{align}
\begin{split}
\braket{K} = & \sum\limits_{q'',q'''} \sum\limits_{m',n'} \frac{1}{2^N}  \sum\limits_{q,q'} \sum\limits_{m,n} \sum_{i} \left \{ \kappa e^{+i\pi (q+\frac{N}{2})} \underbrace{\langle \bar{n}_{-q'}|_{2}n'_{q'''}\rangle_{2}}_{\delta_{\bar{n},n'}\delta{-q',q'''}} \langle n_{q'}|_{1} c_{i1}^{\dagger} |m'_{q''}\rangle_{1} \underbrace{\langle m'_{q''}|_{1} m_{q}\rangle_{1}}_{\delta_{m',m}\delta_{q'',q}} \langle n'_{q'''}|_{2} c_{i2} |\bar{m}_{-q}\rangle_{2} \right. \\
 + & \left. \kappa^{\ast} e^{+i\pi (q''+\frac{N}{2})} \langle \bar{n}_{-q'}|_{2}c_{i2}^{\dagger}|n'_{q'''}\rangle_{2} \underbrace{\langle n_{q'}|_{1} m'_{q''}\rangle_{1}}_{\delta_{n,m'}\delta_{q',q''}} \langle m'_{q''}|_{1}c_{i1} |m_{q}\rangle_{1} \underbrace{\langle n'_{q'''}|_{2}\bar{m}_{-q}\rangle_{2}}_{\delta_{n',\bar{m}}\delta_{q''',-q}}\right \}.
\end{split}
\end{align}
The phase factor $e^{+i\pi (q+\frac{N}{2})}$ arises when we commute $c_{i2}$ across states in system 1 to act on site $i$ of system 2. Summing over Kronecker delta symbols we find
\begin{equation}\label{A4}
\braket{K}= \frac{1}{2^N}  \sum_{q,q'} \sum_{m,n} \sum_{i}  \left(\kappa e^{+i\pi (q+\frac{N}{2})} \langle n_{q'}|_{1} c_{i1}^{\dagger} |m_{q}\rangle_{1} \langle \bar{n}_{-q'}|_{2} c_{i2} |\bar{m}_{-q}\rangle_{2} + \kappa^{\ast} e^{+i\pi (q'+\frac{N}{2})}  \langle \bar{n}_{-q'}|_{2} c_{i2}^{\dagger}|\bar{m}_{-q}\rangle_{2}   \langle n_{q'}|_{1} c_{i1} |m_{q}\rangle_{1} \right).
\end{equation}
We now observe that matrix elements in Eq.\ \eqref{A4} can be evaluated separately for system 1 and 2 which, importantly,  are identical. The matrix elements can thus differ between the systems at most by a phase. We choose this phase so that it  cancels 
the phase factors present  in Eq.\ \eqref{A4}, namely 
\begin{align}
\label{eq:TFDconvention2}
\langle \bar{n}_{-q'}|_{2} c_{i2} |\bar{m}_{-q}\rangle_{2} = -e^{-i\pi (q+\frac{N}{2})} e^{-i\phi} \langle m_{q}|_{1}c_{i1}|n_{q'}\rangle_{1}.
\end{align} 
This corresponds to the following definition of the anti-unitary symmetry
\begin{align}\label{eq:TFDconvention3} \ket{\bar{n}_{-q}}_{2} = \Theta| n_{q}\rangle_{1} = e^{-\eta\frac{i\pi \Gamma}{4}} e^{-iq(\phi-\frac{\pi}{2})} P \ket{n_{q}}_{1}, \end{align}
where $\Gamma = (-1)^{q+\frac{N}{2}}$ is the fermion parity of the SYK eigen-state $\ket{n_{q}}$ and $P = \prod_{i}(c_{i1}^{\dagger} + c_{i1})$ and $\eta =(-1)^{\frac{(N-1)(N+2)}{2}}$ is a sign that depends on total number of fermion in SYK model such that $P^{-1} c_{i1} P = \eta c_{i1}^{\dagger}$. To check this note that the expectation in Eq~(\ref{eq:TFDconvention2}) is only non-zero when $q=q'-1$ and a similar argument holds also for $\kappa^{\ast}$ term when $q=q'+1$ s.t. \begin{align} \langle\bar{n}_{-q'}|_{2} c_{i,2}|\bar{m}_{-q}\rangle_{2} = e^{-i\phi} e^{i\frac{\pi}{2}}e^{-i\eta\frac{\pi}{4}(\Gamma'-\Gamma)}\langle n_{q'}|_{1} P^{-1} c_{i1} P|m_{q}\rangle_{1} \delta_{q,q'-1}\end{align}
where phases are complex conjugated. Noticing $(\Gamma'-\Gamma)$ is $2$ or $-2$ when $q+\frac{N}{2}$ is odd or even, we can replace the phase simply by $-\eta e^{-i\phi} e^{-i\pi (q+\frac{N}{2})}$. Using $P^{-1} c_{i1} P = \eta c_{i1}^{\dagger}$, the above expression becomes \begin{align}-e^{-i\phi} e^{-i\pi (q+\frac{N}{2})} (\langle n_{q'}|_{1} c_{i1}^{\dagger} |m_{q}\rangle_{1})^{\ast} = -e^{-i\phi} e^{-i\pi (q+\frac{N}{2})} \langle m_{q}|_{1} c_{i1} |n_{q'}\rangle_{1}                                                                                    \end{align}
Eq.~(\ref{eq:TFDconvention3}) gives the TFD definition quoted in the main text, Eq.~\eqref{iTFD}, with the required phase factors.

The remaining task is to show that such an  expression for TFD gives the proper expectation value for the tunneling operator. Substituting Eq.\ \eqref{eq:TFDconvention2}  in Eq.~(\ref{eq:TFDconvention1}), the expectation value becomes  
\begin{align} 
\braket{K} = - \frac{\kappa e^{-i\phi} + \kappa^{\ast} e^{i\phi}}{2^{N}} \sum_{q,q'}\sum_{m,n} \sum_i \langle n_{q'}|_{1} c_{i1}^{\dagger} |m_{q}\rangle_{1} \langle m_{q}|_{1} c_{i1} |n_{q'}\rangle_{1},
\end{align} 
where both the states and operators now refer to system $1$. Summing over $|m_{q}\rangle$ and recalling that $\kappa = |\kappa| e^{i \phi}$, we have 
\begin{align}
\braket{K} = - \frac{2|\kappa|}{2^{N}} \sum_{n,q} \sum_i \bra{ n_{q}}_{1} c_{i1}^{\dagger} c_{i1} \ket{n_{q}}_{1} = -\frac{2|\kappa|}{2^N} \sum_{q} (n_{q}) {N \choose n_{q}}  = -|\kappa| N.
\end{align} 
We used the fact that the number of states in charge sector $q$ with fermion number $n_{q}=q+\frac{N}{2}$ is ${N \choose n_{q}}$ and that
\begin{equation}
    \sum\limits_{n_{q}=0}^{N} n_{q} {N \choose n_{q}} = \frac{N}{2}2^{N}.
\end{equation} 

\section{Large $N$ Schwinger-Dyson equations}
\label{app:largeN}
The partition function of our model, in the Euclidean time formalism, is
\begin{align} 
\mathcal{Z} = \int \prod_{i}\prod_{a=1,2} \mathcal{D}c^{\dagger}_{ia} \mathcal{D}c_{ia} \exp\left[ - \int_0^\beta d\tau \left( \sum_{i,a=1,2} c^{\dagger}_{ia}(\tau) \partial_\tau  c_{ia}(\tau)  +  H \right) \right] \end{align}
where
\begin{align} H =& \sum_{i,j,k,l} J_{ijkl} \left( \sum_a c^{\dagger}_{ia} c^{\dagger}_{ja} c_{ka} c_{la} + \alpha \left( c^{\dagger}_{i1} c^{\dagger}_{j2} c_{k1} c_{l2}+ c^{\dagger}_{i2} c^{\dagger}_{j1} c_{k2} c_{l1} \right) \right)
-\mu \sum\limits_{i,a} c^{\dagger}_{ia} c_{ia} + i \kappa \sum_{i} (c^{\dagger}_{i1} c_{i2}-c^{\dagger}_{i2} c_{i1}) 
\end{align} 
and the imaginary-time dependence of the Grassmann variables, $c_{ia}(\tau)$ is implied. In order to perform the disorder average, one must first rewrite the Hamiltonian $H$ in a way which makes its symmetries explicit. In other words, we only want to sum over independent couplings $J_{ijkl}$. The SYK term (diagonal in $a$) becomes
\begin{align}
   \sum_{i,j,k,l} J_{ij;kl} \sum_a c^{\dagger}_{ia} c^{\dagger}_{ja} c_{ka} c_{la} =  4 \sum_{i<j, k<l} J_{ij;kl} \sum_a c^{\dagger}_{ia} c^{\dagger}_{ja} c_{ka} c_{la} 
\end{align}
Similarly, using permutations the $\alpha$ dependent term can be written as 
\begin{align*} 
2\alpha \sum\limits_{i<j,k<l} J_{ij;kl} \left[ c^{\dagger}_{i1} c^{\dagger}_{j2} c_{k1} c_{l2} + c^{\dagger}_{i2} c^{\dagger}_{j1} c_{k1} c_{l2} +c^{\dagger}_{i1} c^{\dagger}_{j2} c_{k2} c_{l1} + c^{\dagger}_{i2} c^{\dagger}_{j1} c_{k2} c_{l1} \right]
\end{align*}

Let us now focus on the interacting part of the action (involving the coupling constants $J_{ijkl}$). We can perform a quenched disorder average to calculate the averaged partition function
\begin{align} 
\overline{\mathcal{Z}_{int}} = \int \mathcal{D}[J,J^*] P(J_{ijkl}) \mathcal{Z}_{int} =  \int \mathcal{D}[c^{\dagger}, c] \int \mathcal{D}[J,J^*] P(J_{ijkl}) \exp \left[ - \left( J_{ijkl}  \phi_{ijkl} + J^{\ast}_{ijkl} \phi_{klij} \right) \right]
\end{align}
where we defined a short-hand notation combining the Hermitian conjugate terms into a single permutation  
\begin{equation}
    \mathcal{D}[J, J^*] \equiv \prod\limits_{\substack{i<j<k<l\\i<k<j<l\\i<k<l<j}} d J_{ij;kl} d J^*_{ij;kl} \quad , \quad \mathcal{D}[c^\dagger, c] \equiv \prod_{i} \prod_{a=1,2} \mathcal{D} c^\dagger_{ia} \mathcal{D} c_{ia}.
\end{equation}
 Here $\phi_{ijkl}$ are four-fermion terms
\begin{equation}
\phi_{ijkl} = \int d\tau \left( 4 \sum_a c^{\dagger}_{i a} c^{\dagger}_{j a} c_{k a} c_{l a}  + 2\alpha \left( c^{\dagger}_{i1} c^{\dagger}_{j2} c_{k1} c_{l2} + c^{\dagger}_{i2} c^{\dagger}_{j1} c_{k1} c_{l2}+ c^{\dagger}_{i1} c^{\dagger}_{j2} c_{k2} c_{l1} + c^{\dagger}_{i2} c^{\dagger}_{j1} c_{k2} c_{l1} \right) \right)
\end{equation}
and $P(J_{ijkl}) = e^{-\frac{|J_{ijkl}|^2}{\sigma^2}}$ is the complex Gaussian distribution with variance $\sigma^2 \equiv \langle |J_{ijkl}|^2\rangle = J^{2}/8N^{3}$. 
Integrating over Gaussian random variables one gets the averaged expression(upto a multiplicative constant)
\begin{align}
    \int d J_{ijkl} d J^*_{ijkl} e^{-\frac{1}{\sigma^2}(J_{ijkl} + \sigma^2 \phi_{klij}) (J_{ijkl} + \sigma^2 \phi_{klij})^{\ast}} e^{\sigma^2 \phi_{ijkl} \phi_{klij}} \equiv e^{\sigma^2 \phi_{ijkl} \phi_{klij}}.
\end{align}
Expressing the averaged partition function with all possible permutations we find
\begin{align}
\overline{\mathcal{Z}_{int}} \equiv \int \mathcal{D}[c^{\dagger}, c] \prod\limits_{\substack{i<j\\k<l}} \exp \left[ \frac{J^{2}}{16 N^{3}} \phi_{ijkl} \phi_{klij} \right]
 = \int \mathcal{D}[c^{\dagger}, c] \exp \left[ \frac{J^{2}}{(4N)^{3}}\sum\limits_{i,j,k,l} \phi_{ijkl} \phi_{klij} \right].
\end{align}
This quenched disorder average is known to be equivalent, for the SYK model, to the more rigorous method of performing the replica trick to average the free energy
\begin{equation}
   \overline{ \ln(\mathcal{Z}) }= \lim\limits_{n\rightarrow0} \frac{1}{n} (\overline{\mathcal{Z}^{n}}-1).
\end{equation} 
The reason is that, for the saddle-point solution of the SYK model, the replica off-diagonal terms can be ignored as they do not contribute to zeroth order in $1/N$. Performing the replica trick with only replica-diagonal terms is formally equivalent to the quenched disorder average.

Combining with the free part of the action, we finally obtain the averaged partition function for the fermions $\overline{\mathcal{Z}} = \int \mathcal{D}[c^{\dagger}, c] e^{-S}$ with the effective action
\begin{align} 
 S &= \int d\tau d\tau' \left( \sum_{i = 1}^N \sum_{a, b} c^{\dagger}_{ia}(\tau')((\partial_\tau -mu)  \delta_{a,b} +i\kappa \epsilon_{a,b}) \delta(\tau - \tau')  c_{ib}(\tau) - \frac{J^{2}}{(4N)^{3}}\sum_{i,j,k,l} \phi_{ijkl} \phi_{klij} \right).
\end{align}

We now integrate out fermions by introducing the averaged Green's functions $G_{ba}(\tau', \tau) = \frac{1}{N} \sum_{i=1}^N \braket{\mathcal{T}  c_{ib} (\tau') c^{\dagger}_{ia}(\tau)}$ through the identity
\begin{equation}
1 \sim \int \mathcal{D}\Sigma \exp  \left( N \int d\tau d\tau' \sum_{a,b} \Sigma_{ab}(\tau, \tau') \left [G_{ba}(\tau', \tau) - \frac{1}{N}\sum_{i=1}^{N} c_{ib}(\tau') c^{\dagger}_{ia}(\tau) \right] \right)
\end{equation}
where the Lagrange multipliers $\Sigma_{ab}(\tau, \tau')$ play the role of fermionic self-energies. After integrating out fermions, the effective $(G,\Sigma)$ action for the averaged Green's functions and self-energies becomes
\begin{eqnarray} 
-\frac{1}{N}S\left[G,\Sigma \right] &= &\ln{\rm Det}((\partial_{\tau} - \mu) \delta_{ab} + i \kappa \epsilon_{ab} - \Sigma_{ab}) + \int d\tau d\tau ' \Big\{\sum_{a,b} \Sigma_{ab}(\tau, \tau') G_{ba}(\tau', \tau) \nonumber +  \frac{J^2}{4} \Big(\sum_{a,b} G_{ab}(\tau, \tau')^{2} G_{ba}(\tau', \tau)^{2} \nonumber \\
&+& 2\alpha \bigl[ G_{11}(\tau , \tau') G_{11}(\tau', \tau) +  G_{22}(\tau , \tau') G_{22}(\tau', \tau) \bigr] \bigl[ G_{12}(\tau', \tau)  G_{21}(\tau, \tau') + G_{21}(\tau' , \tau)  G_{12}(\tau, \tau') \bigr]  \nonumber \\&+& \alpha^{2}\bigl[G_{11}(\tau', \tau) G_{22}(\tau', \tau) G_{11}(\tau, \tau') G_{22}(\tau, \tau') + G_{12}(\tau', \tau) G_{21}(\tau', \tau) G_{11}(\tau, \tau') G_{22}(\tau,\tau') \nonumber \\&+& G_{11}(\tau', \tau) G_{22}(\tau', \tau) G_{21}(\tau,\tau') G_{12}(\tau,\tau') + G_{12}(\tau', \tau) G_{21}(\tau', \tau) G_{21}(\tau, \tau') G_{12}(\tau, \tau') \bigr] \Big) \Big\}
\end{eqnarray}
This expression can be simplified using time translation invariance $G_{ab}(\tau,\tau') = G_{ab}(\tau-\tau')$ and the $R$ symmetry transformation that sends $c_{1} \rightarrow c_{2}$ and $c_{2} \rightarrow -c_{1}$, and implies $G_{11}(\tau) = G_{22}(\tau), \ G_{12}(\tau) = -G_{21}(\tau)$. This leads to 
\begin{align} 
-\frac{1}{N}S\left[G,\Sigma \right] &= \ln{\rm Det}((\partial_{\tau} - \mu) \delta_{ab} + i \kappa \epsilon_{ab} - \Sigma_{ab}) + 2 \beta \int d\tau \Big\{ \Sigma_{11}(\tau) G_{11}(-\tau) + \Sigma_{12}(\tau) G_{21}(-\tau) \nonumber \\ 
&+\frac{J^2}{4} \Big(  \left(1 + \frac{\alpha^2}{2} \right) \bigl[ G^2_{11}(\tau) G^2_{11}(-\tau) + G^2_{12}(\tau) G^2_{12}(-\tau) \bigr] -  4\alpha G_{11}(\tau ) G_{11}(-\tau)  G_{12}(\tau) G_{12}(-\tau)   \nonumber \\
&- \frac{\alpha^{2}}{2} \left[ G^2_{11}(\tau) G^2_{12}(-\tau)  +  G^2_{12}(\tau) G^2_{11}(-\tau) \right] \Big) \Big\},
\end{align}
with a factor of $\beta$ coming from $\int d\tau$ and a factor $2$ coming from adding identical terms. The saddle point equations can be written using ${\delta S[G,\Sigma]/\delta \Sigma} = 0$ and $\delta S[G,\Sigma]/\delta G = 0$.  It is convenient to take the functional derivative with respect to $\Sigma$ in Fourier space using the
convention $$f(\tau) = \frac{1}{\beta} \sum_{\omega_{n}} e^{-i\omega_n \tau} f(\omega_n), \quad f(\omega_n) = \int\limits_{0}^{\beta} d\tau e^{i\omega_n \tau} f(\tau)$$ with $\omega_{n} = {(2n+1)\pi/\beta}$ the Matsubara frequency.
The action with only the $\Sigma(\omega)$ dependent terms thus reads
\begin{align}
 \frac{S}{N} = \ln \mbox{Det} (M) +  2 \sum_{n}\left[\Sigma_{11}(\omega_{n}) G_{11}(\omega_{n})-\Sigma_{12}(\omega_{n}) G_{12}(\omega_{n})\right] + \cdots,  
\end{align}
where the dots represent terms not explicitly containing $\Sigma(\omega_{n})$  and
\begin{equation}
    M= \bigoplus_n \begin{pmatrix}-i\omega_{n}-\mu - \Sigma_{11} & i\kappa - \Sigma_{12} \\ -i\kappa+\Sigma_{12} & -i\omega_{n}-\mu - \Sigma_{11}\end{pmatrix}.
\end{equation}
Finally using ${\delta \ln{\rm Det}(M(\omega_{n}))/\delta \Sigma_{mn}(\omega_{n'})} = 2 (M^{-1})_{nm} \delta(\omega_{n}-\omega_{n'})$ one obtains the saddle point equations \eqref{SDeqn} and \eqref{SDalpha1}  quoted in the main text.

\section{\label{appendix:Conformalsol}Conformal solution for generic $\alpha$}

In this Appendix we examine whether a low-energy scale-invariant solution of the saddle-point equations Eqs.~(\ref{SDeqn},\ref{SDalpha1},\ref{SDalpha2}) for both correlators $G_{11}$ and $G_{12}$ is possible in the gapless phase. We find that this is only possible for the special cases where either $\kappa=0$ or $\alpha=1$, but not for generic parameter choices. 

Our strategy is to assume a power-law ansatz for $G_{11}$ and $G_{12}$ and iterate through the saddle-point equations, in the low-energy limit where the term $-i \omega_n$ can be neglected, to check for consistency. We adopt a power-law ansatz with the same exponent $\nu$ for both correlators. This is necessary since the saddle-point equations involve a sum of squares of the two self energies $\Sigma_{ab}$ -- thus a conformal solution with different exponents for the two correlators can never be self-consistent. Similar to the case of decoupled SYK models, we find that the exponent is enforced to be $\nu = -1/2$. 
Using the imaginary time reflection symmetries valid at charge neutrality $\mu=0$, we thus write the ansatz
\begin{align}\label{C1}
G_{11}(\tau) = b_{11} |\tau|^{-1/2} \mbox{sgn}(\tau) \ , \quad G_{12}(\tau) = i b_{12} |\tau|^{-1/2},
\end{align}
where $b_{11}$ and $b_{12}$ are real numbers. Inserting \eqref{C1} into  the saddle point equations the self-energies become
\begin{align}
\label{selfenergytau}
\Sigma_{11}(\tau) =&  J^{2} \left((1+\frac{\alpha^{2}}{2}) b_{11}^{3} - (2 \alpha - \frac{\alpha^{2}}{2} ) b_{11}b_{12}^2 \right) |\tau|^{-3/2} \ \mbox{sgn}(\tau) = s_{11} |\tau|^{-3/2} \ \mbox{sgn}(\tau), \nonumber\\
\Sigma_{12}(\tau) =&  + i J^{2} \left((1+\frac{\alpha^{2}}{2}) b_{12}^{3} - (2 \alpha - \frac{\alpha^{2}}{2} ) b_{12}b_{11}^2 \right) |\tau|^{-3/2} = i s_{12} |\tau|^{-3/2} ,
\end{align} 
where we implicitly defined the real constants $s_{11}$ and $s_{12}$. Note that at the special point $\alpha=1$ we have a simple relation $s_{11} b_{12} = - s_{12} b_{11}$. Our goal is now to determine the constants $b_{11}$ and $b_{12}$ using the other two saddle-point equations. At $\beta=\infty$, Fourier transforming the first two-point correlator Eq.~\eqref{C1} we get
\begin{align}
G_{11}(i\omega > 0) &= b_{11} \int_{-\infty}^{\infty} d \tau e^{i\omega \tau} |\tau|^{-1/2} \mbox{sgn}(\tau) = 2i b_{11} \int_{0}^{\infty} d\tau \sin(\omega \tau) \tau^{-1/2} = 2i b_{11} \ \omega^{-1/2} \sqrt{ \frac{\pi}{2}}
\end{align}
For negative frequencies the result has an overall negative sign because of the $\sin(\omega \tau)$ function inside the integral. Repeating this calculation for $G_{12}$ and the self energies in Eq.~\eqref{selfenergytau} one obtains 
\begin{align}
  G_{11}(i\omega) &= i \sqrt{2 \pi} b_{11} \ |\omega|^{-1/2} \mbox{sgn}(\omega), \quad
  G_{12}(i\omega) = i \sqrt{2\pi} b_{12} \ |\omega|^{-1/2} \\
  \Sigma_{11}(i\omega) &= 2\sqrt{2 \pi} i s_{11} |\omega|^{1/2} \mathrm{sgn}(\omega) \ , \quad \Sigma_{12}(i\omega) =  - 2\sqrt{2 \pi} i s_{12} |\omega|^{1/2}. \label{C5}
\end{align}
For non-zero $\kappa$, the self-energies given in Eq.\ \eqref{C5} will be consistent with the saddle point equations if we perform a uniform shift and define the modified self-energy
\begin{align}
\tilde{\Sigma}_{12}(i\omega) =  \Sigma_{12}(i \omega) - i \kappa
\end{align}
Here we consider for simplicity the $\mu=0$ limit where the original symmetry of the correlators about $\tau=0$ in Eq.~(\ref{C1}) remains intact. For non-zero $\mu$, one can use a similar redefinition of the self-energy $\tilde{\Sigma}_{11}(i\omega) =  \Sigma_{11}(i \omega) + \mu$ to solve the saddle-point equations in the conformal limit. This choice leads to an asymmetry in $G_{11}(\tau)$ about $\tau=0$ (see Eq.~(\ref{eq:correlator_SYK})) and, as shown in Ref.~\cite{Sachdev2015}, to a twisted $G_{11}(z)$ with a $\mu$-dependent phase factor in the complex frequency $z$ plane. 

Note however that adding a constant frequency shift does not affect the long-time conformal scaling of $\Sigma_{ab}(\tau)$, as it translates to a delta function at early times. The Schwinger-Dyson equations now read
\begin{equation}
\label{eqn:SDeqn}
G_{11}(i\omega) = - \frac{ \tilde{\Sigma}_{11}(i\omega)}{\tilde{\Sigma}^2_{11}(i\omega) + \tilde{\Sigma}^2_{12}(i\omega)} \quad , \quad G_{12}(i\omega) = + \frac{ {\tilde\Sigma_{12}(i\omega) }}{{ \tilde\Sigma^2_{11}(i\omega) + \tilde\Sigma^2_{12}(i\omega)}}
\end{equation}
which lead to
\begin{align}
\label{coeffcondition}
\frac{s_{11}}{s_{12}^2  + s_{11}^2}=  4 \pi b_{11} \quad ,\quad \frac{s_{12}}{s_{12}^2 + s_{11}^2} = - 4 \pi b_{12}.
\end{align}
We thus have two constraints for the two unknown scaling parameters $b_{11}$ and $b_{12}$ which should give us a solution for all values of $\alpha$. However for generic $\alpha$ we find that the only real solution  has $b_{12}=0$ and
\begin{align} b_{11} = \left(\frac{1}{4 \pi J^2 (1+\frac{\alpha^2}{2})}\right)^{1/4}\end{align}
This solution represents decoupled SYK models with no correlations (at the saddle-point level) between the two sides, where the only effect of $\alpha$ was to renormalize the constant $b_{11}$. This solution can thus only represent the $\kappa=0$ limit of our model. This is indeed the numerically obtained solution for $\kappa=0$ and $0 < \alpha < 4$, in the gapless phase. On the other hand, we do not find a conformal solution for $\alpha<0$ or $\alpha > 4$ as the system develops a gap through the symmetry breaking mechanism discussed in the main text.

For non-zero $\kappa$ we always obtain a non-vanishing correlator $G_{12}$ which is inconsistent with the solution above. Thus a conformal solution cannot be found for generic points inside the gapless dome. However, at the SU(2) symmetric point $\alpha=1$ the equations above have additional structure. Using the relation $s_{11} b_{12} = - s_{12} b_{11}$, it is clear that the two equations \eqref{coeffcondition} are now equivalent. We thus have an under-constrained system, which we can solve for
\begin{equation}
    b_{12}^4 = b_{11}^4 - \frac{1}{6\pi J^2}.
\end{equation}
As shown in Fig.~\ref{fig:correlator_alpha}, this relation appears to be satisfied numerically for $\alpha=1$ and small $\kappa$. To fix the value of $b_{11}$, we notice that there is  another constraint coming from the relation
\begin{equation}
    2 i G_{12}(\tau = 0) = \frac{\braket{{K}}}{ \kappa N},
\end{equation}
where ${K}$ is the tunneling operator. For $\alpha=1$ the value of $\braket{K}$ is a good quantum number of the system, because $[{K}, H_{\alpha=1}] = 0$. This is similar to the case of finite chemical potential $\mu$, for which\cite{Sachdev2015}
\begin{equation}
    G_{11}(\tau \rightarrow 0^+) = \frac{1}{2} - \mathcal{Q},
\end{equation}
where $\mathcal{Q}$ is the conserved U(1) charge of the system, and is related to the asymmetry parameter $\cal{E}$ appearing in the low-energy Green's function, Eq.~(\ref{eq:correlator_SYK}) in imaginary time~\cite{SY1993, Sachdev2015}. It is possible to directly compute the value of $\mathcal{Q}$ from the microscopic theory~\cite{Georges2001}, and relate it to the conformal scaling parameter ${\cal E}$ through Eq.~\eqref{eq:Qvsepsilon}. Similarly, here we have at zero temperature
\begin{equation}
   \frac{\braket{K}}{2\kappa N} = i G_{12}(\tau \rightarrow 0) = i \int_{-\infty}^\infty \frac{d\omega}{2\pi} G_{12}(\omega)
\end{equation}
which provides the second constraint allowing to fix $b_{11}$ and $b_{12}$. It is not clear if an analytical result can be obtained for this constraint, as the full form of $G_{12}(\omega)$ at all energies is needed.

\section{Saddle-point equations in real time and frequency}
\label{app:realtime}

In this appendix we show how to analytically continue the imaginary-time saddle point equations \eqref{SDeqn} and (\ref{SDalpha1}-\ref{SDalpha2}), to real time and frequency. The basic scheme is the same as the one described in Refs.\ [\onlinecite{Maldacena2016,Banerjee2017,LH2019,Plugge2020}], but we here summarize the essential steps and results to keep this work self-contained.

\subsection{Analytic continuation of self-energy}
First, let us note a generic U(1)-symmetry conserving self-energy such as the ones appearing in Eqs.~(\ref{SDalpha1}-\ref{SDalpha2}) as
\begin{equation}\label{eq:selfenergy-tau}
\Sigma(\tau) = G_a(\tau)G_b(\tau)G_c(-\tau)~.
\end{equation}
Here $a,b,c$ are arbitrary labels, and prefactors like $J^2$ are omitted.
We first Fourier transform into imaginary frequency
\begin{eqnarray}\label{eq:selfenergy-imag}
\Sigma(i\omega_n) =
\frac{1}{\beta^2} \sum_{n_1,n_2} G_a(i\omega_{n_1})G_b(i\omega_{n_2})
G_c[i(\omega_{n_1}+\omega_{n_2}-\omega_{n})]~.
\end{eqnarray}
Using the spectral representation (Hilbert transform) of the Greens function, $G_x(i\omega_k) = \int d\omega \frac{\rho_x(\omega)}{i\omega_k - \omega}$~, we obtain
\begin{equation}\label{eq:selfenergy-FT'd}
\Sigma(i\omega_n) = \int_{-\infty}^\infty d\omega_{1,2,3}\rho_a(\omega_1)\rho_b(\omega_2)\rho_c(\omega_3)\cdot Y(i\omega_n; \omega_1,\omega_2,\omega_3)
\end{equation}
where, noting the constraint $\omega_{n_3} = \omega_{n_1}+\omega_{n_2}-\omega_{n}$, we have
\begin{equation*}
Y(i\omega_n; \omega_1,\omega_2,\omega_3)= 
\frac{1}{\beta^2}\sum_{n_1,n_2}
\frac{1}{(i\omega_{n_1}-\omega_1)}\frac{1}{(i\omega_{n_2}-\omega_2)}\frac{1}{(i\omega_{n_3}-\omega_3)}~.
\end{equation*}
We now perform the Matsubara summations and use Bose- and Fermi-function identities to evaluate $Y(i\omega_n;\omega_{1,2,3})$.
The $n_2$ sum can be evaluated by defining 
$\Omega = i\omega_n-i\omega_{n_1} + \omega_3$ to obtain
\begin{eqnarray}\label{eq:Y-A}
\frac{1}{\beta}\sum_{n_2}
\frac{1}{(i\omega_{n_2}-\omega_2)}\frac{-1}{(\Omega-i\omega_{n_2})} e^{i\omega_{n_2}0^+}
= \frac{n_F(\Omega)-n_F(\omega_2)}{\Omega - \omega_2}
= \frac{n_F(\omega_3)-n_F(\omega_2)}{i\omega_n-i\omega_{n_1} + \omega_3 -\omega_2}~.\qquad\qquad
\end{eqnarray}
In the last step we used the fact that the imaginary part of $\Omega$ is a multiple of $2\pi T$, hence $n_F(\Omega) = n_F(\omega_3)$.
To evaluate the $n_1$ sum we define $\tilde{\Omega} = i\omega_n +\omega_3 -\omega_2$ and get
\begin{eqnarray}\label{eq:Y-B}
\frac{1}{\beta}\sum_{n_1}
\frac{1}{(i\omega_{n_1}-\omega_1)}\frac{1}{(\tilde{\Omega}-i\omega_{n_1})} e^{i\omega_{n_1}0^+}
= \frac{n_F(\omega_1)-n_F(\tilde{\Omega})}{\tilde{\Omega} - \omega_1}
= \frac{n_F(\omega_1)+n_B(\omega_3-\omega_2)}{i\omega_n + \omega_3 -\omega_2 - \omega_1}~.\qquad\qquad
\end{eqnarray}
Here we used that $\tilde{\Omega}$ is a fermionic Matsubara frequency, hence $n_F(\tilde{\Omega}) =-n_B(\omega_3-\omega_2)$.
To get $Y$, we now take the product of the numerator in Eq.~\eqref{eq:Y-A} and the expression Eq.~\eqref{eq:Y-B}. The product of both numerators simplifies to
\begin{eqnarray}
[n_F(\omega_1)+n_B(\omega_3-\omega_2)]
[n_F(\omega_3)-n_F(\omega_2)]
= n_F(\omega_1)n_F(\omega_3) - n_F(\omega_1)n_F(\omega_2)
- n_F(-\omega_2)n_F(\omega_3)
\notag\\\notag
= -n_F(\omega_1)n_F(\omega_2)n_F(-\omega_3) - n_F(-\omega_1)n_F(-\omega_2)n_F(\omega_3)~.
\end{eqnarray}
In the last step we inserted identities $1 = n_F(\omega_j) + n_F(-\omega_j)$ to obtain a more symmetric expression.
Finally we get
\begin{equation}\label{eq:Y-C}
Y(i\omega_n; \omega_1,\omega_2,\omega_3)= 
-\frac{n_F(\omega_1)n_F(\omega_2)n_F(-\omega_3) +n_F(-\omega_1)n_F(-\omega_2)n_F(\omega_3)}{i\omega_n -\omega_1 -\omega_2 +\omega_3}~.
\end{equation}
Note the frequency-symmetric form $\omega_j \to -\omega_j$ of the numerator.
Inspecting Eq.~\eqref{eq:Y-C} and the self-energy in Eq.~\eqref{eq:selfenergy-FT'd}, the remaining imaginary frequency $i\omega_n$ now appears only in the denominator of $Y$. We hence can analytically continue $i\omega_n \to \omega +i\eta$ to obtain the retarded self-energy from Eq.~\eqref{eq:selfenergy-FT'd} with $Y(\omega+i\eta; \omega_1,\omega_2, \omega_3)$ given in Eq.~\eqref{eq:Y-C}.
We then use the identity $\frac{1}{\bar{\Omega}+ i\eta} = -i\int_0^\infty dt e^{i(\bar{\Omega}+i\eta)t}$ with $\bar{\Omega} = \omega - \omega_1- \omega_2 + \omega_3$ to obtain
\begin{eqnarray}\label{eq:selfenergy-retarded}
\Sigma^{\mathrm{ret}}(\omega) = i\int_0^\infty dt \int_{-\infty}^\infty d\omega_{1,2,3}
e^{i(\omega + i\eta -\omega_1 -\omega_2 +\omega_3)t}
\rho_a(\omega_1)\rho_b(\omega_2)\rho_c(\omega_3)
\left[
n_F(\omega_1)n_F(\omega_2)n_F(-\omega_3) +(\omega_j \leftrightarrow -\omega_j) \right]~.\qquad
\end{eqnarray}
This allows us to perform the three frequency integrals, and finally we obtain
\begin{equation}\label{eq:selfenergy-general}
\Sigma^{\mathrm{ret}}(\omega) = i \int_0^\infty dt e^{i(\omega+i\eta)t}
\left[ n_{a}^{++}n_{b}^{++}n_{c}^{-+} + n_{a}^{+-}n_{b}^{+-}n_{c}^{--} \right].
\end{equation}
Here we defined the ``time-dependent occupations''
\begin{equation}\label{eq:nss}
n_x^{ss'}(t) = \int_{-\infty}^\infty d\omega \rho_x(s\omega)n_F(s'\omega) e^{-i\omega t}~,
\end{equation}
that can be calculated directly from the spectral function, and hence from the retarded Greens functions.
As we will note below, analytically continuing the Greens functions is essentially trivial, and hence the expressions (\ref{eq:selfenergy-general}-\ref{eq:nss}) are convenient for the numerical solution of the saddle-point equations in real time and frequency~\cite{Maldacena2016,LH2019,Plugge2020}.

\subsection{Application to the coupled complex SYK model}

The retarded self-energies $\Sigma_{11,12}^\mathrm{ret}(\omega)$ are obtained from the analytical continuation of Eqs.~(\ref{SDalpha1}-\ref{SDalpha2}), according to the recipe outlined above. It is useful to simplify these expressions further by taking some of the observed spectral symmetries into account.
The general form reads $\Sigma_{x=11,12}^\mathrm{ret}(\omega) = -i J^2 \int_0^\infty dt e^{i(\omega+i\eta)t} K_{x}(t)$ with
\begin{eqnarray}\label{eq:K11t}
K_{11}(t) =
(1+\frac12\alpha^2) ~ [(n_{11}^{++})^2 n_{11}^{-+} + (n_{11}^{+-})^2 n_{11}^{--} ]
- 2\alpha ~ [ n_{11}^{++}n_{12}^{++}n_{12}^{-+} + n_{11}^{+-}n_{12}^{+-}n_{12}^{--} ] \qquad
\\ \notag
- \frac12\alpha^2 ~ [ (n_{12}^{++})^2 n_{11}^{-+} + (n_{12}^{+-})^2 n_{11}^{--} ]
\end{eqnarray}
and
\begin{eqnarray}\label{eq:K12t}
-K_{12}(t) =
(1+\frac12\alpha^2) ~ [(n_{12}^{++})^2 n_{12}^{-+} + (n_{12}^{+-})^2 n_{12}^{--} ]
- 2\alpha ~ [ n_{12}^{++}n_{11}^{++}n_{11}^{-+} + n_{12}^{+-}n_{11}^{+-}n_{11}^{--} ] 
\qquad
\\ \notag
- \frac12\alpha^2 ~ [ (n_{11}^{++})^2 n_{12}^{-+} + (n_{11}^{+-})^2 n_{12}^{--} ]
\end{eqnarray}
Note that, up to an overall minus sign, $K_{11}(t)$ and $K_{12}(t)$ are directly related by replacing $11\leftrightarrow 12$ everywhere.
To make further progress, note the simple form of the  non-interacting retarded Green's functions
\begin{equation}\label{eq:invGFnonint}
[g_{11}(\omega)]^{-1} = \omega-\mu + i\eta~,~~~
[g_{12}(\omega)]^{-1} = -i\kappa~.
\end{equation}
Following the convention for spectral functions in Ref.~\cite{Zhang2020}, and using $G_{21}^{\mathrm{ret}}(\omega) = -G_{12}^{\mathrm{ret}}(\omega)$, we obtain
\begin{equation}\label{eq:spectralfunct}
\rho_{11}(\omega) = -\frac{1}{\pi}{\rm Im} G_{11}^{\mathrm{ret}}(\omega),~~
\rho_{12}(\omega) = \frac{i}{\pi}{\rm Re} G_{12}^{\mathrm{ret}}(\omega).
\end{equation}
 In plots in the main text, cf. Figs. \ref{fig:revivals} and \ref{fig:Gpm-kmu}, when referring to $\rho_{12}$ we implicitly consider ${\rm Im}[\rho_{12}(\omega)]$ with the above definition.
One can simplify the above expressions for $K_{11,12}$ by using properties of the spectral functions $\rho_{11,12}$. First $\rho_{11}$ ($\rho_{12}$) is purely real (imaginary), and for zero chemical potential $\mu = 0$ they also have a definite frequency parity:
\begin{eqnarray}\notag
\rho_{11}(\omega) &=& [\rho_{11}(\omega)]^\ast~~~,~~
\rho_{11}(\omega) = \rho_{11}(-\omega)~,
\\\notag
\rho_{12}(\omega) &=& -[\rho_{12}(\omega)]^\ast~,~~
\rho_{12}(\omega) = -\rho_{12}(-\omega)~.
\end{eqnarray}
Using these properties, one can express all $n_{11}^{\pm\pm}$ and $n_{12}^{\pm\pm}$ by a single $n_{11}$ and $n_{12}$. We note
\begin{eqnarray}
n_{11} \equiv n_{11}^{++}
\stackrel{\omega}{=} +n_{11}^{-+}
\stackrel{\mathrm{Re}}{=} +[n_{11}^{+-}]^\ast
\stackrel{\omega}{=} +[n_{11}^{--}]^\ast
~,\quad
\\
n_{12} \equiv n_{12}^{++}
\stackrel{\omega}{=} -n_{12}^{-+}
\stackrel{\mathrm{Im}}{=} +[n_{12}^{+-}]^\ast
\stackrel{\omega}{=} -[n_{12}^{--}]^\ast
~,\quad
\end{eqnarray}
where at $\omega$ we used the frequency parity, and $\mathrm{Re}/\mathrm{Im}$ means we used the real/complex-valuedness of $\rho_{11/12}$.
Then
\begin{eqnarray}\label{eq:K11t-simple}
K_{11}(t) &=&
(2+\alpha^2)\mathrm{Re}[n_{11}^3]
+ (4\alpha -\alpha^2)\mathrm{Re}[n_{11}n_{12}^2]
~,~~\qquad
\\ \label{eq:K12t-simple}
K_{12}(t) &=&
(2+\alpha^2)\mathrm{Re}[n_{12}^3]
+ (4\alpha -\alpha^2)\mathrm{Re}[n_{12}n_{11}^2]
~.~~~\qquad
\end{eqnarray}
This simplified version makes apparent the symmetry of self-energies under exchange $11\leftrightarrow 12$. It also suggests that there is a second `decoupling point'' at $\alpha = 4$, similar to the situation at $\alpha = 0$. However the SYK interaction strength at this point is enhanced to an effective $J_{\alpha=4} = 3J$. We further discuss this around Eq.~\eqref{eq:sigmas-alpha4} and in Fig.~\ref{gapscalingmu}.

The above SD equations are solved numerically by repeated self-energy evaluations and re-insertion into the Dyson equation~\eqref{SDeqn} until a fixed point solution is found.
As the starting point we use
$g_{11}(\omega)$ and $g_{12}(\omega)$ in Eq.~\eqref{eq:invGFnonint} with the initial value $\kappa_0 = 0, \kappa$ or $\kappa^{2/3}$. We then check that the same solution is obtained independent of the starting point and the iteration parameters.

\end{document}